\documentclass[aps,pra,floatfix,superscriptaddress,twocolumn,showpacs,10pt]{revtex4-1}

\usepackage{amssymb, amsmath,color,mciteplus,graphicx,subfigure}
\usepackage{calc,epsfig,epstopdf,color,mciteplus,bm,mathrsfs}
\usepackage{times}
\usepackage{float}
\usepackage{multirow}
\usepackage{lipsum}

\begin{document}

\title{Composite Short-path Nonadiabatic Holonomic Quantum Gates}

\author{Yan Liang}\email{These two authors contributed equally to this work.}
\affiliation{Guangdong Provincial Key Laboratory of Quantum Engineering and Quantum Materials,
and School of Physics\\ and Telecommunication Engineering, South China Normal University, Guangzhou 510006, China}

\author{Pu Shen}\email{These two authors contributed equally to this work.}
\affiliation{Guangdong Provincial Key Laboratory of Quantum Engineering and Quantum Materials,
and School of Physics\\ and Telecommunication Engineering, South China Normal University, Guangzhou 510006, China}

\author{Tao Chen}
\affiliation{Guangdong Provincial Key Laboratory of Quantum Engineering and Quantum Materials,
and School of Physics\\ and Telecommunication Engineering, South China Normal University, Guangzhou 510006, China}

\author{Zheng-Yuan Xue}\email{zyxue83@163.com}
\affiliation{Guangdong Provincial Key Laboratory of Quantum Engineering and Quantum Materials,
and School of Physics\\ and Telecommunication Engineering, South China Normal University, Guangzhou 510006, China}

\affiliation{Guangdong-Hong Kong Joint Laboratory of Quantum Matter, Frontier Research Institute for Physics,\\ South China Normal University, Guangzhou 510006, China}

\date{\today}

\begin{abstract}
Nonadiabatic holonomic quantum computation (NHQC)
has attracted significant attention due to its fast evolution
and the geometric nature induced resilience to local noises. However, its long operation time and complex physical implementation make it hard to surpass the dynamical scheme, and thus hindering its wide application. Here, we present to implement NHQC with the shortest path under some conditions, through the inverse Hamiltonian engineering technique, which  posseses higher fidelity and stronger robustness than previous NHQC schemes. Meanwhile, the gate performance in our scheme can be further improved by using the proposed composite dynamical decoupling pulses, which can efficiently improve both the gate fidelity and robustness, making our scheme outperform the optimal dynamical scheme in certain parameters range. Remarkably, our scheme can be readily implemented with Rydberg atoms, and a simplified implementation of the controlled-not gate in the Rydberg blockade regime can be achieved. Therefore, our scheme represents a promising progress towards future fault-tolerant quantum computation in atomic systems.

\end{abstract}

\maketitle

\section{INTRODUCTION}

Manipulating quantum states in a robust way is the necessary condition to realize large-scale quantum computing \cite{Nielson}, and has been attracted  much attention  in various  systems, such as cavity quantum electrodynamics (QED) \cite{cavity}, trapped ions \cite{ions}, neutral atoms in optical lattices \cite{atom1,atom2}, etc. The Rydberg atom is one of the promising physical systems due to its excellent atomic properties, including strong and long-range interaction, giant polarizability, and long lifetime \cite{ Saffman2010}.

One of the main obstacles of manipulating quantum systems is how to implement them in a robust way. Holonomic quantum computation \cite{hqc}  is one of the well-known strategies  for improving the gate robustness due to its geometric properties
\cite{Solinas2004, Solinas2012, Johansson2012}. However, early proposals are based on the adiabatic evolution \cite{hqc, JP99, Duan2001}, which require a long gate time, and thus leads to unacceptable decoherence-induced error. To break this limitation and shorten the needed evolution time, the nonadiabatic holonomic quantum computation (NHQC) is proposed \cite{ Sjoqvist2012, xu2012}, which  
becomes a promising method to realize quantum computation.  The early NHQC schemes are implemented based on the resonance three-level model \cite{ Sjoqvist2012, xu2012} and been experimentally demonstrated in various quantum systems \cite{Abdumalikov2013, long2013, Duan2014, SAC2014,SDanilin2018}. But for the realization of an arbitrary single-qubit gate, it needs to concatenate two separate cycles, which will increase the decoherence-induced error. To remove this obstacle, researchers have come up with the improved approaches that enable the realization of the arbitrary single-qubit gate through a single-loop evolution \cite{xu2015,ESjoqvist2016,Herterich2016, hong2018}, which have also been experimentally verified  \cite{Sekiguchi2017, long2017, zhou2017, sun2018, NI2018, peng2019}.

Although the improved approaches above reduce the evolution time to a certain extent, it is still useful to further shorten the holonomic gate time to reduce the effect of decoherence, as it is still much longer than a typical gate from conventional dynamical evolution. This is a nontrivial work as two conditions must be met to achieve NHQC: cyclic evolution condition and parallel transport condition, and thus the schemes of NHQC have strict restrictions on the evolution time.
To further shorten the gate time, combining with the time-optimal technology \cite{Carlini2012,Carlini2013,wamg2015,geng2016}, the universal holonomic gates with  minimum time can be obtained \cite{liuarxiv,chen2020,jiln2021,shenp2021} by solving the quantum brachistochrone equation, which has been experimentally verified \cite{yu2020, sunfw2021}. However, due to this additional constraint, the geometry phase obtained there is of an unconventional nature \cite{zhu2003, du2006}. Another method of getting faster holonomic quantum gates is achieved via shortening the evolution path \cite{xu2018,zhao2020}. In addition, the method of noncyclic holonomic quantum computation breaks the limits of cyclic evolution, thus accelerating the evolution process \cite{SaiLi2020}.
Besides the gate-time consideration, the pulse-shaping technique is also applied in  NHQC schemes \cite{xugf2014, liubj2019, Lisai2020, Lisai2021, liubj2021} with experimental demonstrations \cite{yudp2019, YS2019, aimz2020, aimz2021, sunfwprappl2021, xuy2021}, mainly to strengthen the gate robustness.


Here, we demonstrate how to realize the shortest-path NHQC (SNHQC) via the inverse Hamiltonian engineering technique \cite{Kang2018,Odelin2019} on  three-level $\Lambda$ quantum systems. Under the set conditions, the SNHQC scheme possesses the shortest evolution path with ultrahigh gate fidelity. Remarkably, the gate performance can be further improved by utilizing the proposed composite dynamical decoupling pulse, termed as CSNHQC here. Interestingly, the population of the excited state will decrease with the increase of the composite pulse sequence in our SNHQC, and thus improve both the gate fidelity and robustness. This is distinct from the conventional NHQC schemes \cite{xugf2014, liubj2019, Lisai2020, Lisai2021, liubj2021, yudp2019, YS2019, aimz2020, aimz2021, sunfwprappl2021, xuy2021} with dynamical decoupling pulse and pulse shaping, where the gate robustness is obtained at the cost of decreasing the gate fidelity.
In addition, we compare our CSNHQC scheme with the conventional dynamical scheme, and our scheme performs  better in certain parameters ranges. Finally, we present the implementation of our scheme with Rydberg atoms in the Rydberg blockade regime \cite{Saffman2010}, and show its better gate performance. Therefore,  our scheme represents a promising progress towards the fault-tolerant quantum computation in the atomic system.

\section{UNIVERSAL SINGLE-QUBIT HOLONOMIC GATES}
In this section, we first derive the Hamiltonian that can realize the holonomic quantum gate via inverse engineering in subsection A. We then design the single-qubit gates of our SNHQC scheme in subsection B. Finally, we discuss the robustness of single-qubit gates from the SNHQC scheme and compare them with the previous NHQC one in subsection C.

\subsection{Inverse engineering of Hamiltonian}
{We first consider a set of complete basis vectors $|\Psi_k(t)\rangle$, $k=1,2,...,L$, satisfying the time-dependent Schr\"{o}dinger equation under a Hamiltonian that is to be determined}
\begin{eqnarray}
\label{schrodinger}
H(t)|\Psi_k(t)\rangle=i|\dot{\Psi}_k(t)\rangle.
\end{eqnarray}
There is a unitary evolution operator that can drive the initial state to the final state, i.e., $|\Psi_k(t)\rangle=U(t)|\Psi_k(0)\rangle$, which  can be generally written as
\begin{eqnarray}
\label{2}
U(t) =\!\!\!\sum^L_{k=1}|\Psi_{k}(t)\rangle\langle \Psi_{k}(0)|.
\end{eqnarray}
{Then, the corresponding Hamiltonian can be inversely obtained from the assumed dynamics, i.e.,}
\begin{eqnarray}
\label{3}
H(t) =i\dot{U} (t)U^{\dag}(t)=\!\!i\!\sum^L_{k=1}|\dot{\Psi}_{k}(t)\rangle\langle \Psi_{k}(t)|.
\end{eqnarray}
{Different approaches depend on different choices of the orthogonal basis functions $|\Psi_{k}(t)\rangle$ \cite{Odelin2019}. For example, the invariant-based engineering,  $|\Psi_{k}(t)\rangle$ can be constructed by the eigenstates of the invariant of an assumed Hamiltonian. More generally, $|\Psi_{k}(t)\rangle$ can be convenient functions according to one's need.} In terms of our goal, constructing the gates for NHQC, there are two conditions that we have to follow \cite{Sjoqvist2012, xu2012}: the cyclic and parallel transport conditions. According to these two conditions,  {we choose a set of time-dependent auxiliary vectors $\{|\mu_k(t)\rangle \}_{k=1}^{L+1}$ with $|\mu_k(\tau)\rangle =|\mu_k(0)\rangle$ to denote a set of bases in the $L$+1-dimensional Hilbert space,} which do not need to satisfy the Schr\"{o}dinger equation. Here, $\tau$ is the period of evolution. We then set the evolution states in the $L$+1-dimensional quantum system as
\begin{eqnarray}
\label{4}
&&|\Psi_{k}(t)\rangle=\!\sum_{i=1}^{L}C_{ik}(t)|\mu_{i}(t)\rangle ,k=1,2,....L \notag,\\
&&|\Psi_{L+1}(t)\rangle=e^{i\zeta (t)} |\mu_{L+1}(t)\rangle,
\end{eqnarray}
where the coefficient $C_{ik}(t)$ is a matrix element  of a $L\times L$ matrix $C(t)= \mathcal{T} e^{i\int_{0}^{t}A(t^{'})dt^{'}}$ with $C_{ik}(0)=\delta_{ik}$ and $\mathcal{T}$ being the time-ordering operator; $A_{ij}(t)=i\langle\mu_{i}(t)|\dot{\mu}_{j}(t)\rangle$ and $\zeta (t)$ is a time-dependent real function with $\zeta (0)=0$.

Thus, in the subspace $\{|\Psi_{k}(t)\rangle\}_{k=1}^{L}$, the basis functions $|\Psi_{k}(t)\rangle$ satisfy the cyclic condition $\sum_{k}^{L}|\Psi_{k}(\tau)\rangle\langle\Psi_{k}(\tau)|=\sum_{k}^{L} |\Psi_{k}(0)\rangle\langle\Psi_{k}(0)|$, and the parallel transport condition $\langle\Psi_{k}(t)|\dot{\Psi}_{l}(t)\rangle=0$ $(k,l=1,...L)$. Hence, when we choose the subspace $S_L(0)={\rm Span}$$\{|\Psi_{k}(0)\rangle=|\mu_{k}(0)\rangle\}_{k=1}^{L}$ as the computation space of NHQC, {after a period of cyclic evolution, the evolution operator acting on the subspace $S_L(0)$ can be written as $U(\tau)=C(\tau)=\mathcal{T}e^{i\int_{0}^{\tau}A(t )dt }$,} which is a holonomic gate acting on the $L$-dimensional subspace $S_L(0)$. Therefore, by substituting Eq. (\ref{4}) into Eq. (\ref{3}), the Hamiltonian can be expressed by the auxiliary vectors as \cite{zhao2020}
\begin{eqnarray}
\label{5}
&&H(t)=\left [i\sum_{i=1}^{L}\langle \mu_{i}(t)| \dot{\mu}_{L+1}(t)\rangle |\mu _{i}(t)\rangle\langle \mu_{L+1}(t)|+\rm{H.c.}\right ]\notag\\
&&+\left [i \langle \mu_{L+1}(t)| \dot{\mu}_{L+1}(t)\rangle-\dot{\zeta}(t)\right ]|\mu _{L+1}(t)\rangle\langle \mu_{L+1}(t)|,
\end{eqnarray}
which can be used to construct  nonadiabatic holonomic gates.

\subsection{Arbitrary single-qubit gate of SNHQC}

\begin{figure}[tb]
  \centering
  \includegraphics[width=0.9\linewidth]{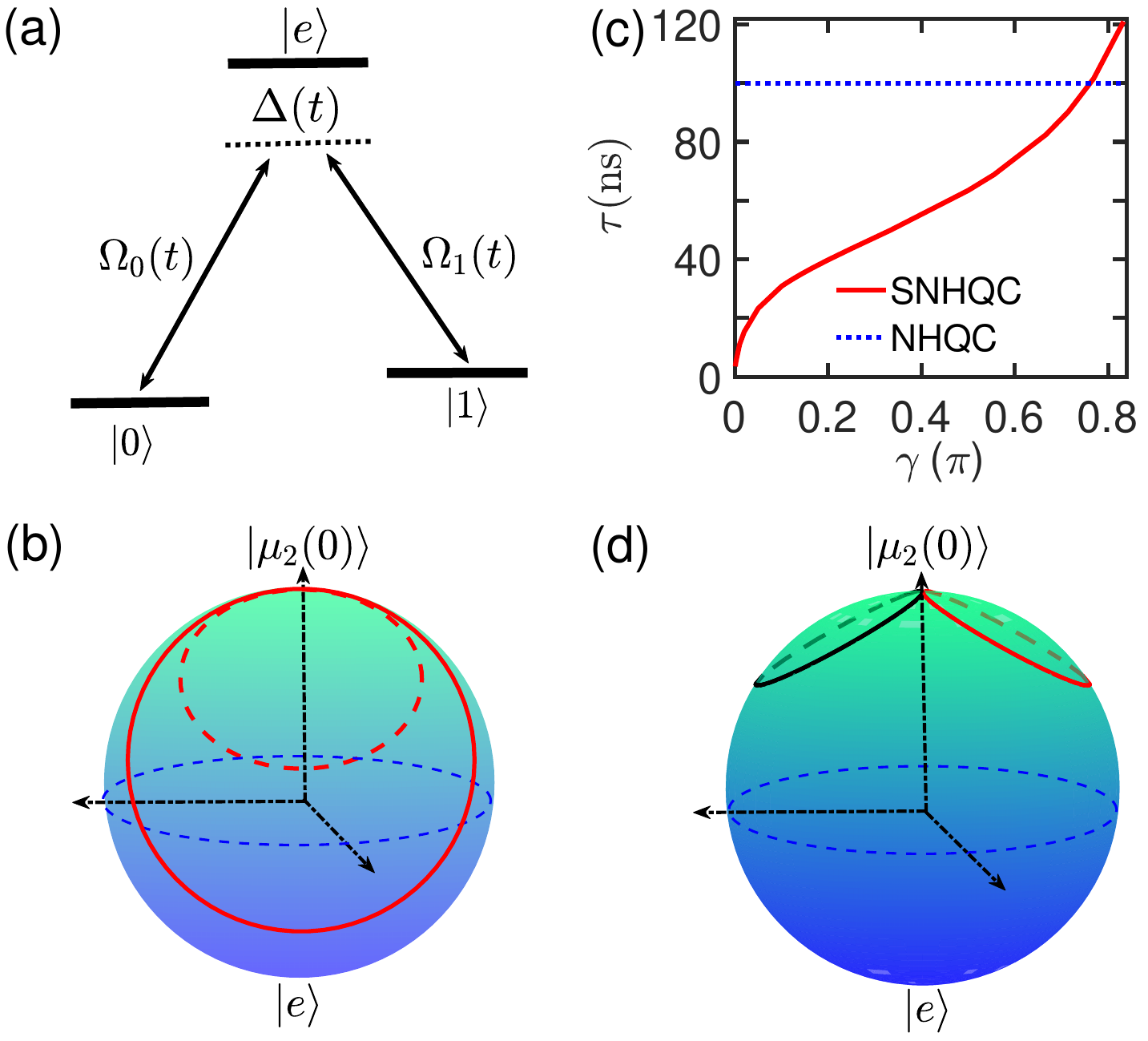}
\caption{(a) Schematic energy levels of a three-level $\Lambda$ system.
(b) Evolution paths of different gates. The dashed line denotes the evolution path of $T$ gate, and the solid line denotes the evolution path of $S$ gate and $\sqrt{\rm H}$ gate. (c) The $Z$-axis-rotation gate time of the SNHQC scheme and NHQC scheme as a function of rotation angles, with the time-dependent pulse shape of NHQC being $\Omega(t) = \Omega_{\rm m} \sin^2({\pi t/\tau})$. (d) The evolution path for the optimized $T$ gate  by using the simplest composite dynamical decoupling pulse.  }\label{fig1}
\end{figure}

We now illustrate the realization of arbitrary single-qubit gate with SNHQC. A three-level $\Lambda$ system is considered as shown in Fig. \ref{fig1}(a), where the  two low-energy levels $|0\rangle$ and $|1\rangle$ are served as our qubit states and a high excited state $|e\rangle$ as the auxiliary state. We define the auxiliary vectors as
\begin{eqnarray}
\label{6}
|\mu_{1}(t)\rangle=&\cos&\frac{\theta}{2}|0\rangle+\sin\frac{\theta}{2}e^{i\varphi}|1\rangle \notag,\\
|\mu_{2}(t)\rangle=&\cos&\frac{\alpha(t)}{2}\left(\sin\frac{\theta}{2}e^{-i\varphi}|0\rangle -\cos\frac{\theta}{2}|1\rangle\right ) \notag\\
&+&\sin\frac{\alpha(t)}{2} e^{i\beta(t)} |e\rangle \notag,\\
|\mu_{3}(t)\rangle=&\sin&\frac{\alpha(t)}{2} e^{-i\beta(t)}\left(\sin\frac{\theta}{2}e^{-i\varphi}|0\rangle-\cos\frac{\theta}{2}|1\rangle\right) \notag\\
&-&\cos\frac{\alpha(t)}{2}|e\rangle ,
\end{eqnarray}
where $\theta, \varphi$ are time-independent parameters, and $\alpha(t),\beta(t)$ denote the time-dependent polar angle and azimuthal angle of a spherical coordinate system with $\alpha(\tau)=\alpha(0)=0$. It is obvious that the subspace
$S_{L}(t)={\rm Span}\{|\mu_{1}(t)\rangle,|\mu_{2}(t)\rangle\}$ undergoes a cyclic evolution at the final time $\tau$. Therefore, we can regard the initial space $S_{L}(0)={\rm Span}\{|\mu_{1}(0)\rangle,|\mu_{2}(0)\rangle\}={\rm Span}\{|0\rangle,|1\rangle\}$ as the computational space.

Meanwhile, by substituting these auxiliary vectors in Eq. (\ref{6}) to Eq. (\ref{5}), we can obtain the following Hamiltonian
\begin{eqnarray}
\label{7}
H(t)=&\bigtriangleup&(t)|e\rangle\langle e|+\{\Omega_{0}(t)e^{-i[\beta(t)+\chi(t)+\varphi]}|0\rangle\langle e| \notag\\
&+&\Omega_{1}(t)e^{-i[\beta(t)+\chi(t)+\pi]}|1\rangle\langle e|+\rm{H.c.}\}.
\end{eqnarray}
It represents that the three-level $\Lambda$ system is driven by two laser fields with Rabi frequencies $\Omega_{0}(t)=\Omega(t)\sin(\theta/2)$ and $\Omega_{1}(t)=\Omega(t)\cos(\theta/2)$, in a two-photon resonant way with a common detuning being $\triangle(t)=-\dot{\beta}(t)\left[1+\cos\alpha(t)\right]$, as shown in Fig. \ref{fig1}(a). The other parameter constraints are
\begin{subequations}\label{para}
\begin{eqnarray}
\Omega(t)=\frac{1}{2}\sqrt{\left[\dot{\beta}(t)\sin\alpha(t)\right]^{2}+\dot{\alpha}^{2}(t)}, \end{eqnarray}
\begin{eqnarray}
\chi(t)=\arctan\left\{\dot{\alpha}(t)\big{/}\left[\dot{\beta}(t)\sin\alpha(t)\right]\right\}.
\end{eqnarray}
\end{subequations}
Besides,
$\dot{\zeta}(t)=\dot{\beta}(t)[3+\cos\alpha(t)]/2$ has to be met to avoid the direct coupling between $|0\rangle$ and $|1\rangle$ states. Then, the Hamiltonian in Eq. (\ref{7}) can be expressed by the auxiliary vectors as
\begin{eqnarray}
\label{8}
H(t)&\!=\!&\Delta(t)|e\rangle\langle e|\!+\!\left\{\Omega(t)e^{-i[\beta(t)\!+\!\chi(t)]}|\mu_{2}(0)\rangle\langle e|\!+\!\rm{H.c.}\right\}. \notag \\
\end{eqnarray}
{In our scheme, the detuning $\Delta(t)$ is time dependent, which  can be realized by the time-dependent manipulation of the frequency of the optical Raman beams \cite{PHLeung2018, CFiggatt2019, LandsmanKA2019}.}

Governed by the Hamiltonian in Eq. (\ref{8}), the unitary operator acting on the computational subspace after a cyclic evolution is
\begin{eqnarray}
\label{9}
U_1(\tau)&=& |\mu_{1}(0)\rangle\langle\mu_{1}(0)|+e^{-i\gamma}|\mu_{2}(0)\rangle\langle\mu_{2}(0)|\notag\\
&=&\rm {exp}\left(i\frac{\gamma}{2}\textbf{n}\cdot\bm{\sigma}\right ),
\end{eqnarray}
where $\textbf{n}=(\sin\theta\cos\varphi,\sin\theta\sin\varphi,\cos\theta)$ is the rotation axis and $\bm{\sigma}$ is the Pauli vector of the computational basis $\{|0\rangle, |1\rangle\}$. The evolution operator denotes a rotation operation around the axis $\textbf{n}$ by an angle
\begin{eqnarray}
\label{10}
\gamma= \frac{1}{2}\int_{0}^{\tau}\dot{\beta}(t)[1-\cos\alpha(t)]dt.
\end{eqnarray}
Since $[\alpha(t),\beta(t)]$ denotes a point on the Bloch sphere, when the quantum system evolves from 0 to $\tau$, the track of $[\alpha(t),\beta(t)]$ is a closed path $C$ in the unit sphere. From a geometric point of view, the angle $\gamma$ can be recast as $\gamma= \oint_{C}[1-\cos\alpha(t)]/2d\beta$. This result leads to the interesting observation that, $\gamma$ is half of the solid angle enclosed by path $C$. This result, in turn, implies that the geometric phase depends only on the certain global properties of the path and thus is robust against local fluctuations. In particular, for a same $\gamma$, different choices of $\alpha(t)$ and $\beta(t)$ determine the different paths of the geometric evolution. Therefore, there have innumerable path options for implementing a certain nonadiabatic holonomic gate, e.g., the orange-slice-shaped loop of the previous scheme \cite{hong2018}, the three-step path  \cite{zhao2020}, etc. However, the shortest path for implementing the nonadiabatic holonomic gate has not been studied in detail.

We then turn to generate the nonadiabatic holonomic gates with the shortest path, i.e., a circle on the Bloch sphere, which we term as SNHQC, and the gate robustness is also discussed. Due to the constraint of $\alpha(0)=\alpha(\tau)=0$, all the  single-qubit holonomic gates have to start from the north pole of the Bloch sphere, and back to the north pole at the final time. In this case, we provide a general set of parameter forms of $\alpha(t)$ and $\beta(t)$ to construct the circle path, i.e.,
\begin{eqnarray} \label{11}
\beta(t)&=&\beta_{0}+\pi\sin^{2}\left(\frac{\pi t}{2\tau}\right),\notag\\
\alpha(t)&=&2\arctan\{\ell\sin[\beta(t)-\beta_{0}]\},
\end{eqnarray}
where $\beta_0$ is the initial value of the azimuthal angle $\beta(t)$, which can be arbitrary, and for the sake of simplicity, we choose $\beta_{0}\!=\!0$; $\tau$ denotes the evolution period of single-qubit gate; $\ell=\sqrt{2\pi \gamma-\gamma^2}/(\pi-\gamma)$, which means  $\gamma\neq\pi$.

Using the evolution operator of Eq. (\ref{9}), we can construct arbitrary single-qubit nonadiabatic holonomic gate. In the following, we focus on three representative single-qubit gates: $S$ gate, $T$ gate, and square root Hadamard $(\sqrt{\rm H})$ gate, which can be constructed by choosing $(\theta, \varphi,\gamma)=(0, 0,  \pi/2)$, $(\theta, \varphi,\gamma)=(0, 0, \pi/4)$, and $(\theta, \varphi,\gamma)=(\pi/4, 0,  \pi/2)$, respectively. Note that the Hadamard ($H$) gate in our scheme cannot be constructed directly as $\gamma\neq\pi$ here, and we present the result of  $\sqrt{\rm H}$ gate, two of which can be composite together to construct an $H$ gate, see Section \ref{Hgate} for its detailed construction and performance. The corresponding evolution paths are shown in the Bloch sphere in Fig. \ref{fig1}(b), where the north pole is denoted as $|\mu_{2}(0)\rangle$ and the south pole is $|e\rangle$. We can see that, the evolution paths of $S$ gate and $\sqrt{\rm H}$ gate are the same circles (solid line) as they possess the same angle of $\gamma_{_{\rm S,\sqrt{\rm H}}} =\pi/2$. The evolution path of $T$ gate (dashed line) is shorter than $S$ gate and $\sqrt{\rm H}$ gate since it possesses a smaller rotation angle $\gamma_{_{\rm T}} =\pi/4$. Actually, the smaller angle of $\gamma$, the shorter circle evolution path. The $Z$-axis-rotation gate time of S-NHQC and conventional NHQC \cite{hong2018} as a function of rotation angles $\gamma$ are shown in Fig. \ref{fig1}(c), where the time-dependent pulse shape of  NHQC is $\Omega(t) = \Omega_{\rm m} \sin^2({\pi t/\tau})$ with $\Omega_{\rm m}=2\pi\times 10$ MHz, and the pulse shape of SNHQC is obtained from  Eqs. (\ref{para}) and (\ref{11}) with the same maximum value. It shows that the evolution time of the NHQC scheme remains 100 ns no matter whether the rotation angles $\gamma$ are big or small, while the evolution time of the SNHQC scheme increases with the increase of the rotation angle $\gamma$. When $\gamma$ is smaller than $0.76 \pi$, the evolution time of the SNHQC scheme is shorter than the NHQC one.

\begin{figure}[t]
  \centering
  \includegraphics[width=0.95\linewidth]{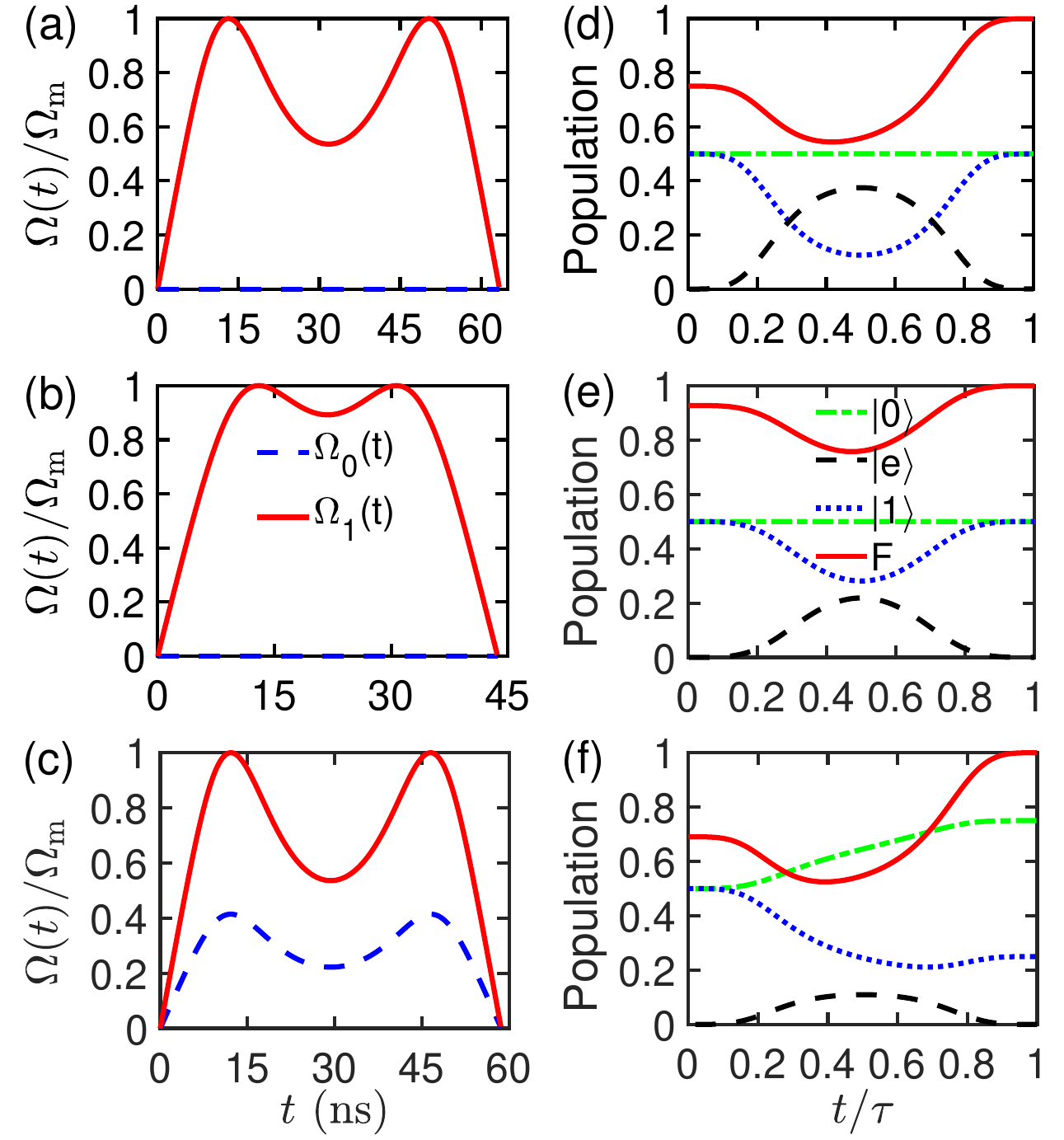}
\caption{ {Time dependence of Rabi frequencies for (a) $S$ gate, (b) $T$ gate, and (c) $\sqrt{\rm H}$ gate, where solid-red and dashed-blue lines
denote $\Omega_1(t)$ and $\Omega_0(t)$, respectively. The state population and gate-fidelity dynamics of (d) $S$ gate, (e) $T$ gate, and (f) $\sqrt{\rm H}$ gate.}}\label{fig2}
\end{figure}

\subsection{Gate performance } \label{performance}
 We further proceed to evaluate the performance of single-qubit gates by using the Lindblad master equation of
\begin{eqnarray}
\label{EqMaster}
\dot\rho&=&-i[H(t), \rho]+\frac {1} {2}\sum_{j=-,z,q}\Gamma_{j}L(\sigma_{j}),
\end{eqnarray}
where $\rho$ is the density matrix of the quantum system and $L(A)=2A\rho A-A^{\dag}A\rho-\rho A^{\dag}A $ is the Lindbladian operator with $\sigma_-=|0\rangle\langle e|+|1\rangle\langle e|$, $\sigma_z=|e\rangle\langle e|-|1\rangle\langle1|-|0\rangle\langle0|$ and $\sigma_{q}=|0\rangle\langle 1|$; $\Gamma_-$ and $\Gamma_z$ represent the decay and dephasing rates, respectively; $\Gamma_{q}$ is the decay rate from $|1\rangle$ to $|0\rangle$. In Figs. \ref{fig2}(a)-(c), we plot the shapes of the Rabi frequency for $S$, $T$, and  $\sqrt{\rm H}$ gates, respectively. Here we choose the maximum value of the Rabi frequency as $\Omega_{\rm m}=2\pi\times10$ MHz, according to typical experimental requirements, which results in gate time of $\tau_{_{\rm S}}=63.45\  \rm{ns}$, $\tau_{_{\rm T}}=43.67 \ \rm{ ns}$ and $\tau_{_{\sqrt{\rm H}}}= 58.62 \ \rm{ns}$.
The state populations are depicted in Figs. \ref{fig2}(d)-(f), with an initial state of $|\psi_0\rangle=(|0\rangle+|1\rangle)/\sqrt{2}$, for $S$ gate, $T$ gate, and $\sqrt{\rm H}$ gate, respectively.  {To fully evaluate the performance of the implemented gates, we also plot the gate-fidelity dynamics with the definition of $F= \langle\psi_f|\rho|\psi_f\rangle $ \cite{hong2018}, where $|\psi_f\rangle$ is the ideal state. The gate fidelities are numerically obtained as the average of 1600 input states of $|\psi'_0\rangle=\cos\theta_1|0\rangle+{\rm exp}(i\phi)\sin\theta_1|1\rangle$} with $\theta_1$ and $\phi$ being uniformly distributed over $[0, 2\pi]$, which are as high as $F_{\rm S}=99.97\%$, $F_{\rm T}=99.99\%$, and $F_{\sqrt{\rm H}}=99.97\%$. The results of gate-fidelity dynamics are also shown in Figs. \ref{fig2}(d)-(f) for $S$, $T$, and $\sqrt{\rm H}$ gates, respectively. Here we have set the decoherence rates of qubits as $\Gamma_-=2\pi\times3$ kHz, $\Gamma_z=\Gamma_0/100$ and $\Gamma_{q}=0$.

\begin{figure}[tbp]
  \centering
\includegraphics[width=\linewidth]{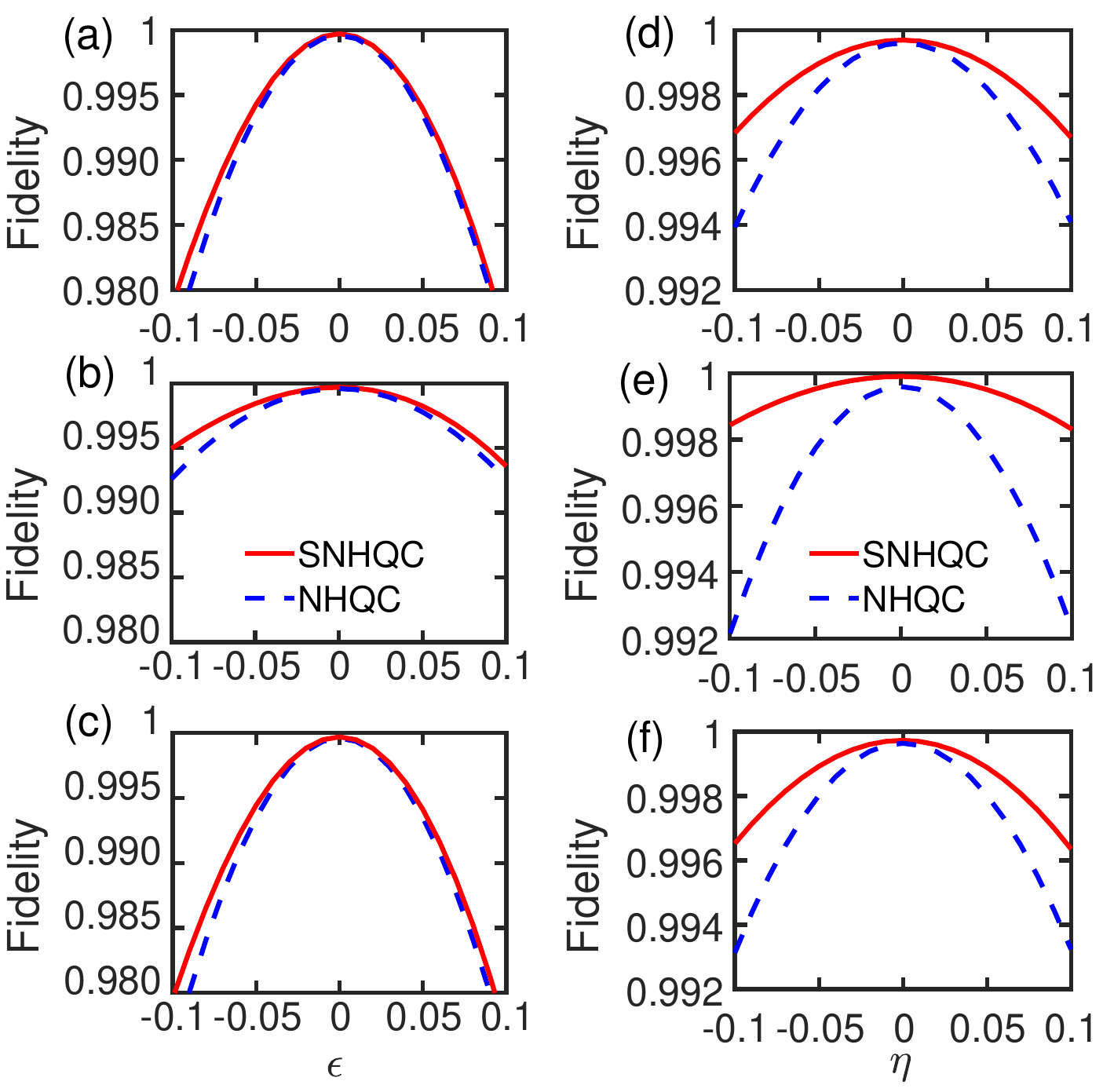}
\caption{ Fidelity of (a) $S$ gate, (b) $T$ gate, and (c) $\sqrt{\rm H}$ gate with respect to the systematic Rabi error. Fidelity of (d) $S$ gate, (e) $T$ gate, and (f) $\sqrt{\rm H}$ gate with respect to the frequency drift error. The solid-red and dashed-blue lines denote the results from SNHQC and NHQC schemes, respectively.}\label{fig3}
\end{figure}

We next turn to consider the case where the error exists, i.e., discuss the robustness of the implemented holonomic quantum gates. To show our construction is more robust than previous single-loop NHQC schemes \cite{hong2018}, we test the robustness with respect to the systematic error. When the implementation is disturbed by noises and/or errors, it can be described by the Hamiltonian of
\begin{eqnarray}
\label{13}
H_{\epsilon,\eta}(t)&\!=\!&\left\{[\Omega(t)+\epsilon\Omega(t)]e^{-i[\beta(t)\!+\!\chi(t)]}|\mu_{2}(0)\rangle\langle e|\!+\!\rm{H.c.}\right\}\notag \\
&+&\![\Delta(t)+\eta\Omega(t)] |e\rangle\langle e|.
\end{eqnarray}
Here we introduce the time-dependent Rabi and  frequency drift errors as $\epsilon\Omega(t)$ and $\eta\Omega(t)$, with the  time-independent error fractions   $\{\epsilon, \eta\} \in [-0.1,0.1]$. The comparison results for gate robustness are shown in Fig. \ref{fig3}, where the decoherence effect is also included as above. Figures \ref{fig3}(a)-(c) show the gate robustness against the systematic Rabi error for $S$ gate, $T$ gate, and $\sqrt{\rm H}$ gate, respectively. Figures \ref{fig3}(d)-(f) correspond to the gate robustness with respect to the frequency drift error for $S$ gate, $T$ gate, and $\sqrt{\rm H}$ gate, respectively. These results clearly show that, our scheme is more robust than  previous single-loop NHQC scheme within the whole considered error range for both the systematic Rabi and frequency drift errors. In particular, our scheme has more advantages in gate robustness against the frequency drift error, which is one of the main concerned errors in solid-state quantum system.

\begin{figure}[t]
  \centering
  \includegraphics[width=\linewidth]{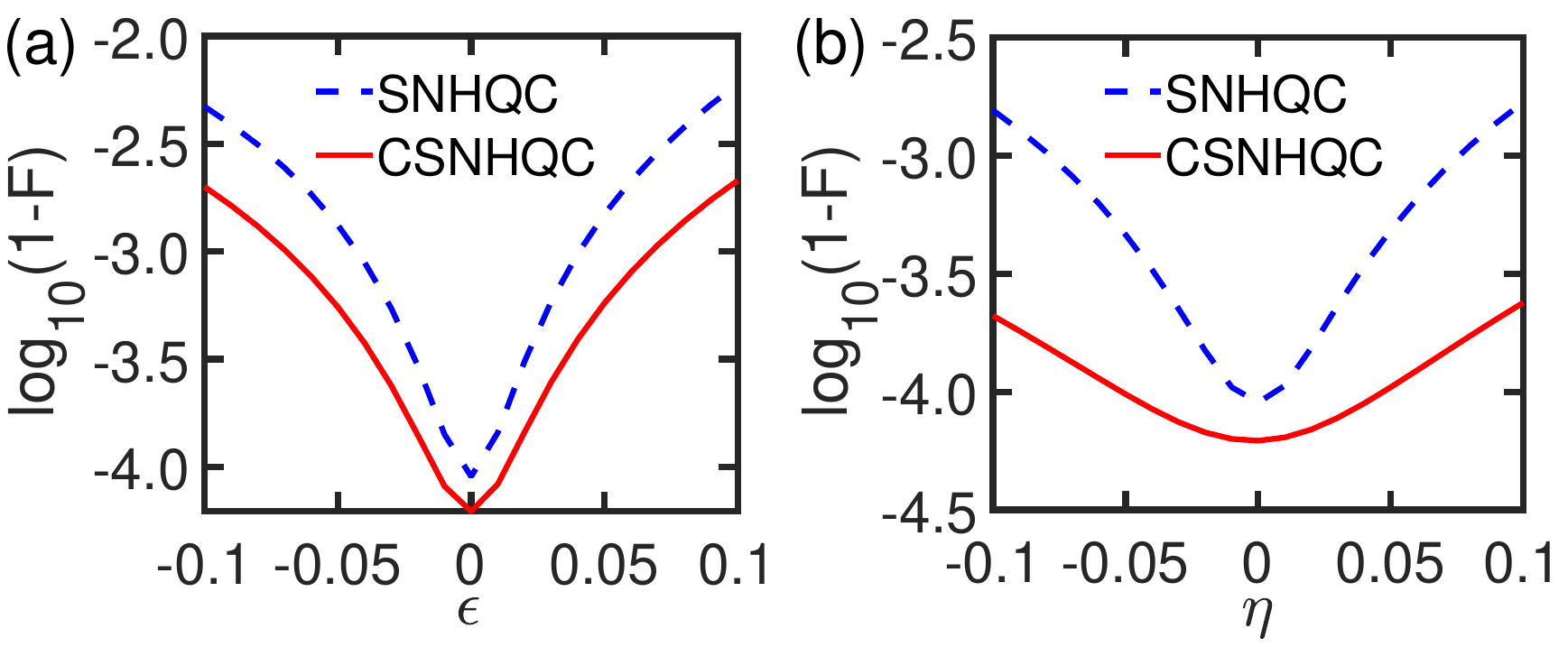}
  \caption{ Performance for the $T$ gate with the composite dynamical decoupling pulse. The gate infidelity with respect to (a) the Rabi error and (b) frequency drift errors. The solid-red and dashed-blue lines denote the CSNHQC and SNHQC schemes, respectively.  }\label{fig4}
\end{figure}

\section{OPTIMIZATION} \label{Hgate}
{Decoherence caused by the interaction between the quantum system and its environment is one of the main barriers to the realization of high-fidelity quantum gates. Dynamical decoupling \cite{Viola1999} provides an efficient way to mitigate the decoherence-induced error via reversing the evolution of the quantum system at specific times with control pulses. Here, we further show that our SNHQC scheme can be optimized by using composite dynamical decoupling pulses, which we term as CSNHQC. Remarkably, we can achieve the decoupling effect  in Ref. \cite{Viola1999} without additional control fields, and thus simplify its realization. Although the composite pulse will lead to a longer evolution time, the CSNHQC scheme can greatly reduce the population of the excited state, thereby reducing the decoherence-induced error and improving the obtained gate fidelity, which is a distinct merit of our scheme. As a Benefit of this reduction, we can synthesize any large-angle rotation gate, in a way that is more robust than the NHQC scheme, by using the optimization of composite dynamical decoupling pulse that needs only small-angle rotation, and thus solve the problem that large-angle ($\gamma>0.76\pi$) operations are susceptible to decoherence due to the long evolution time.} This also makes it possible for the CSNHQC scheme to surpass the dynamical gate (DG) scheme. In the following, we implement the CSNHQC scheme in detail and compare it with the DG scheme.

\subsection{The optimization of $T$ gate}
\begin{figure}[tbp]
  \centering
  \includegraphics[width=0.9\linewidth]{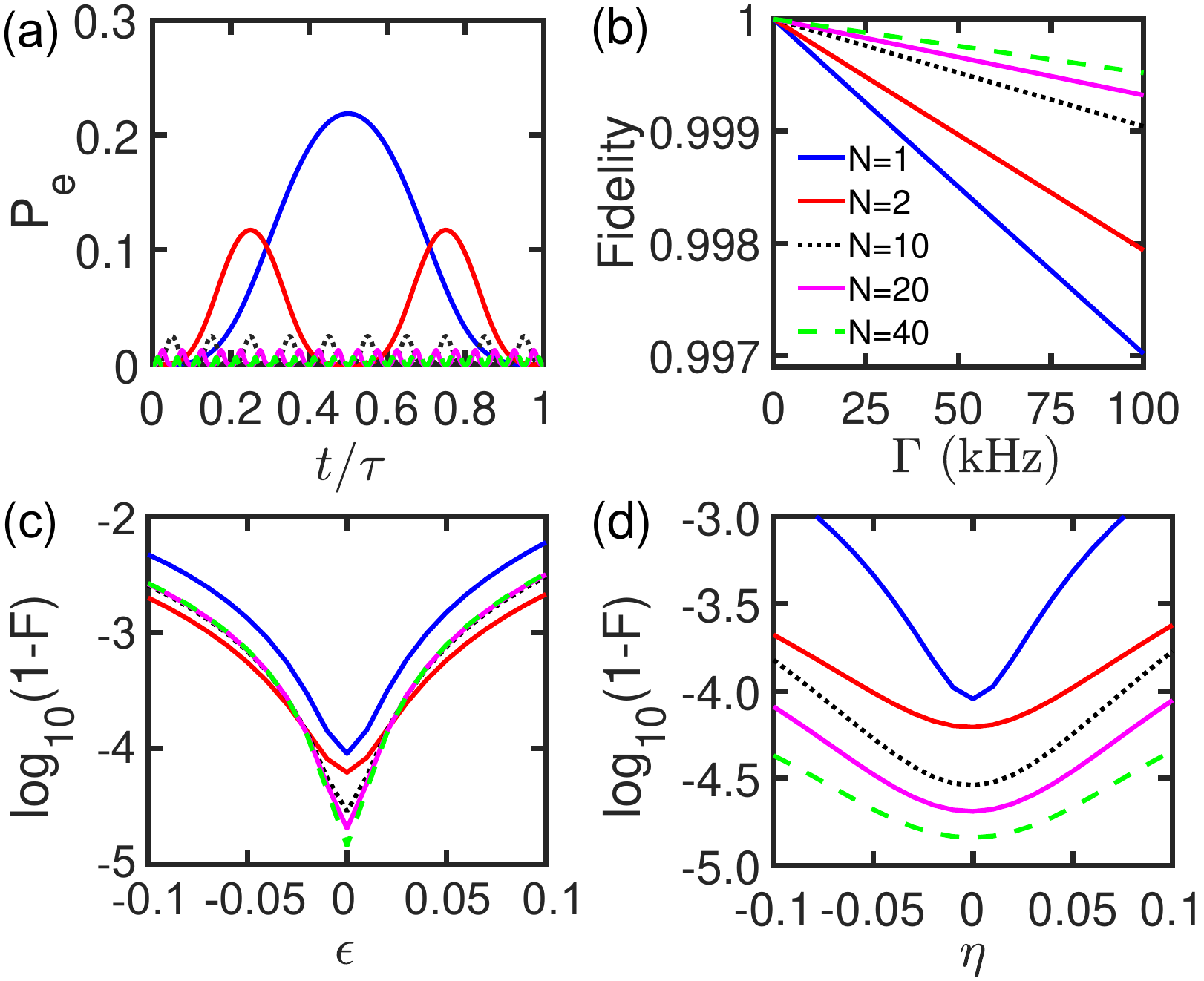}
\caption{ The performance for the optimized $T$ gate with different composite dynamical decoupling pulse sequences and the comparison with the unoptimized SNHQC scheme ( $N$=1 ). (a) The excited-state population. (b) The gate fidelity with respect to decoherence. The infidelity of gate with respect to the (c) systematic Rabi error and (d) frequency drift error.}
\label{Fig5}
\end{figure}

For a train of $N$ ($N$ is a positive integer) pulse sequences, the evolution operator $U$ can be expressed as $U$ = $U_N...U_3U_2U_1$.
We first consider the evolution operator $U(\theta, \varphi, \gamma)$ of the CSNHQC gate in the simplest case with $N$ = 2. During the first interval $t$ $\in$ $[0, \tau]$, we set the Hamiltonian as the form in Eq. (\ref{8}) with $\beta_0$ = 0. The corresponding evolution operator is $U_1(\theta, \varphi, \gamma/2)$. For the second interval $t\in[\tau, 2\tau]$, the Hamiltonian is still in the form of Eq. (\ref{8}) but $\beta_0$ = $\pi$, and thus the corresponding evolution operator is $U_2(\theta, \varphi, \gamma/2)$. Hence, the CSNHQC gate can be written as $U(\theta, \varphi, \gamma)=U_2(\theta, \varphi, \gamma/2) U_1(\theta, \varphi, \gamma/2)$. For the $H$ gate, it can be obtained by $\rm H=U_2(\pi/4, 0, \pi/2) U_1(\pi/4, 0, \pi/2)$.

We further take $T$ gate for example, where $\theta=0, \varphi=0, \gamma=\pi/4$, to detail our optimization. Note that, other gates have similar properties, and thus will not be shown here. Figure \ref{fig1}(d) depicts the evolution path of the optimized $T$ gate  with two composite pulse sequences. We can see that the evolution path is two symmetrical circles at the north pole, which we use to offset the decoherence effect during the evolution process. We also show the infidelities of $T$ gate with respect to the systematic Rabi error in Fig. \ref{fig4}(a) and the frequency drift error in Fig. \ref{fig4}(b), respectively, where we take the decoherence rates as $\Gamma_-=2\pi \times 3$ kHz, $\Gamma_z=\Gamma_-/100$, and $\Gamma_{q}=0$. These results show that, in the entire considered parameter range, the CSNHQC scheme can not only combat decoherence but also enhance the robustness of gates to the system Rabi error and frequency drift error.

Remarkably, considering a train of $N$ pulses ($N\geq2$ and to be even number), where $\beta_0$ equals 0 and $\pi$ for all the odd and even pulses, respectively. By this setting, the evolution operator for all the odd and even pulses will be the same as $U_1(\theta, \varphi, \gamma/2)$ and $U_2(\theta, \varphi, \gamma/2)$, respectively. We find the following three merits of our CSNHQC scheme. Firstly, as shown in Fig. \ref{Fig5}(a), the larger $N$ is, the smaller population of the excited state $|e\rangle$ will be. It is similar to the case of the large-detuned dynamical scheme in the three-level $\Lambda$ system, where the population of the excited state $|e\rangle$ decreases with the increase of detuning.  {Secondly, the capability of combating decoherence enhances as $N$ increases, as shown in Fig. \ref{Fig5}(b). Therefore, we can achieve the dynamical decoupling from the environment through a large $N$. It also indicates that, we can improve the fidelity of large-angle rotation operations through the optimization of the CSNHQC scheme, despite the evolution time of large-angle rotation gate does longer than that of the NHQC scheme. Finally, the gate infidelity in the presence of the frequency drift error becomes smaller with the increase of $N$. As shown in Fig. \ref{Fig5}(d), the gate infidelity can be smaller than $10^{-4}$ when $N$ is larger than 20.} However, for the systematic Rabi error, the improvement will be quickly saturated, and the best gate robustness is achieved with $N=2$, as shown in Fig. \ref{Fig5}(c).

\subsection{Comparison with the dynamical scheme}
The optimized results above are obtained by considering only partial decoherence effect, i.e., $\Gamma_-$ and $\Gamma_z$ in Eq. (\ref{EqMaster}),  with $\Gamma_{q}=0$. However, it is clear that the total evolution time will increase with the increase of $N$, so that the effect of $\Gamma_{q}$ can no longer be neglected, despite that, it is very small compared with the other two. Therefore, we need to confirm the optimal value of $N$, which makes the gate most robust to decoherence, in the presence of $\Gamma_{q}$. Moreover, although the composite pulse scheme possesses a similar effect with the DG scheme with large-detuning, we find that the CSNHQC scheme can surpass the DG one, where both schemes are compared in terms of their corresponding best performance  under the same decoherence rates.

\begin{table}
	\centering

	\begin{tabular}{|c|c|c|c|}
        \hline
		\multirow{4}{*}{Decoherence} & Case 1 & Case 2 & Case 3 \\
		\cline{1-4}
		~  & $\Gamma_{-}=\Omega_{\rm m}/2000$ & $\Gamma_{-}=\Omega_{\rm m}/2000$ & $\Gamma_{-}=\Omega_{\rm m}/100$  \\
		
		~ & $\Gamma_{z}= \Gamma_{-}/100$ & $\Gamma_{z}= \Gamma_{-}/100$ & $\Gamma_{z}= \Gamma_{-}/100$ \\
        ~ &  $\Gamma_{q}= \Gamma_{-}/100$ &  $\Gamma_{q}= \Gamma_{-}/10$ &  $\Gamma_{q}= \Gamma_{-}/10$\\
		\hline
        \multirow{2}*{DG} & $\triangle_1=66\ \tilde{\Omega}_{1}$ & $\triangle_1=32\ \tilde{\Omega}_{1}$ & $\triangle_1=17\ \tilde{\Omega}_{1}$\\
         ~ & $F=99.93\%$& $F=99.79\%$& $F=98.37\%$\\
         \hline
         \multirow{2}*{CSNHQC} & $N=160$ & $N=10$ & $N=10$\\

         ~ & $F=99.99\%$& $F=99.97\%$& $F=98.65\%$\\
         \hline
	\end{tabular}
\caption{ The best performance of the $H$ gate from the CSNHQC and DG schemes, under the different decoherence rates. }
\label{Table1}
\end{table}

\begin{figure}[tbp]
  \centering
  \includegraphics[width= \linewidth]{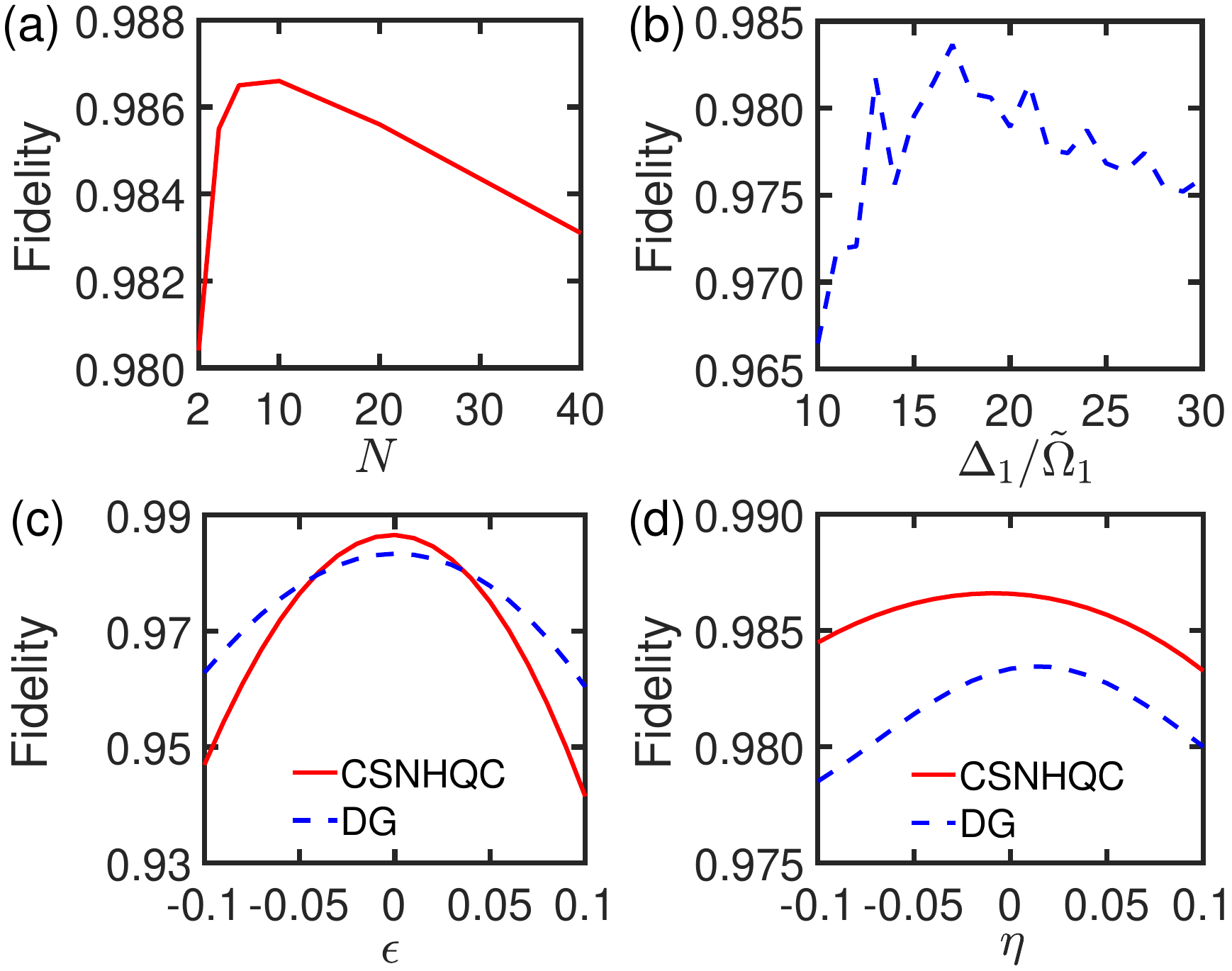}
  \caption{ The $H$ gate performance for case 3. The gate fidelity of  the (a) CSNHQC  and (b) DG schemes with respect  to the pulse sequence number $N$ and the detuning, respectively. The  robustness  with respect to the (c) systematic Rabi error and (d) the frequency drift error, where the solid-red and dashed-blue  lines denote the CSNHQC and DG  schemes, respectively.}
\label{Fig6}
\end{figure}
We take Hadamard gate ($H$ gate) as an example here. Since the $H$ gate can be obtained in one step by the DG method, while the CSNHQC scheme requires to use two $\sqrt{\rm H}$ gates to synthesize an $H$ gate, i.e., $H=U_2(\theta, \varphi, \gamma/2) U_1(\theta, \varphi, \gamma/2)$. To be more specific, the H gate is the easiest one from the DG scheme but the hardest one in our scheme. Therefore, if the $H$ gate performance of our scheme can be better than the DG one, it can strongly support the conclusion that our scheme is better than the DG one.
The dynamical $H$ gate is governed by the following Hamiltonian
\begin{eqnarray}
\label{dynamical}
\tilde{H}(t)&=&(1+\epsilon)\left(\tilde{\Omega}_0|e\rangle\langle0|e^{-i\Delta_{0}t} +\tilde{\Omega}_1|e\rangle\langle1|e^{-i\Delta_{1}t}+\rm{H.c.}\right)\notag \\
&+&\eta\tilde{\Omega}_0|e\rangle\langle e|,
\end{eqnarray}
with $\tilde{\Omega}_0=\left[\int^{\tau}_0\Omega_0(t)dt+\int^{\tau}_0\Omega_1(t)dt\right]/(2\tau)$ equaling the average of Rabi frequency of the CSNHQC scheme. The rest of the parameters satisfy $\Delta_{1}=j\tilde{\Omega}_1 (j>0)$,
$\Delta_{0}=  \tilde{\Omega}^2_{0}\Delta_{1}/ \tilde{\Omega}^2_{1} $ and $\Delta_{0}-\Delta_{1}= \tilde{\Omega}_0\tilde{\Omega}_1(\Delta_{0}+\Delta_{1})/(\Delta_{0}\Delta_{1})$. $\epsilon$ and $\eta$ represent the systematic Rabi error rate and the frequency drift error rate, respectively. The construction of the dynamical $H$ gate is presented in detail in Appendix \ref{appendixA}.

The comparison results are listed in Table \ref{Table1}. When we set the decoherence rates as
$\Gamma_-=\Omega_{\rm m}/2000, \Gamma_z=\Gamma_{q}=\Gamma_{-}/100$, i.e., case 1 in Table \ref{Table1}, the best performance of the DG scheme appears at $\Delta_{1}=66\ \tilde{\Omega}_{1}$, and the CSNHQC scheme can surpass the DG one with $N=160$. When we keep $\Gamma_-$ and $\Gamma_z$ unchanged and increase the decay rate between $|0\rangle$ and $|1\rangle$ to $\Gamma_{q}=\Gamma_{-}/10$, case 2 in Table \ref{Table1}, the best performance of the DG scheme appears at $\Delta_{1}=32 \ \tilde{\Omega}_{1}$, and the CSNHQC scheme can surpass the DG scheme when $N=10$. In the last set of data, case 3 in Table \ref{Table1}, we increase all the decoherence rates simultaneously to $\Gamma_-=\Omega_{\rm m}/100, \Gamma_z=\Gamma_{0}/100, \Gamma_{q}=\Gamma_{-}/10$. In this case, the best performance of the DG scheme appears at $\Delta_{1}=17 \ \tilde{\Omega}_{1}$, and the CSNHQC scheme can surpass the DG scheme at $N=10$. That is, when the decoherence rate $\Gamma_{q}$ increases, our CSNHQC scheme can surpass the DG scheme with fewer pulse sequences. In addition, for different quantum systems, one can use our strategy to find optimal results for both schemes, and for the certain parameter ranges the worst gate performance of our scheme can surpass that of the best gate from the dynamical scheme.

To compare the $H$-gate robustness of our CSNHQC scheme with the DG scheme, we consider the systematic Rabi error and the frequency drift error under the decoherence rates in case 3 above. We first search the optimal pulse sequence $N$ of the CSNHQC scheme and the optimal detuning of the DG scheme. Figures \ref{Fig6}(a) and \ref{Fig6}(b) point out that, the optimal pulse sequence $N$ of CSNHQC scheme is 10, and the optimal detuning of DG scheme is $\Delta_{1}=17\ \tilde{\Omega}_{1}$. With the optimal value of $N$ and $\Delta_{1}$, we plot the fidelities with respect to both the systematic Rabi and frequency drift errors in Figs. \ref{Fig6}(c) and \ref{Fig6}(d), respectively. As shown in Fig. \ref{Fig6}(d), our CSNHQC scheme has obvious advantages in terms of robustness against the frequency drift error, but its performance for the systematic Rabi error is not so good, as shown in Fig. \ref{Fig6}(c). One reason is that the $H$ gate cannot be constructed directly in our CSNHQC scheme, as $\gamma=\pi$ is forbidden here. Another reason is that the composite scheme does not work that well for the systematic Rabi error, the improvement effect will be saturated when $N$ = 2, as mentioned before. This justifies that the $H$ gate in our scheme is the hardest, and thus the performance of which is the worst.

\begin{figure}[tbp]
  \centering
  \includegraphics[width= \linewidth]{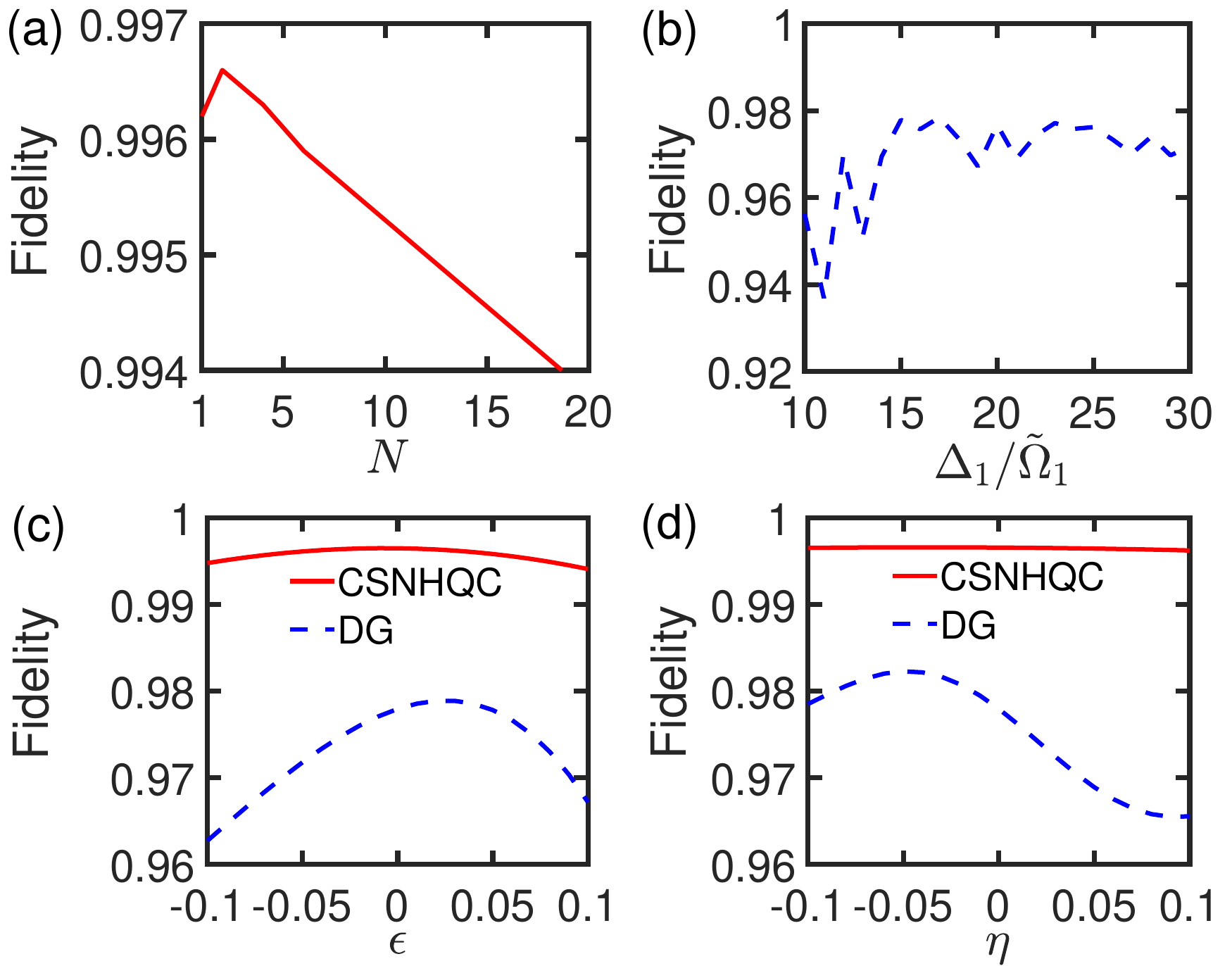}
  \caption{ Numerics for $T$-gate performance, where the decoherence rates are the same as in case 3. The gate fidelity of  the (a) CSNHQC and (b)  DG schemes with respect  to the pulse sequence number $N$ and the detuning, respectively. The robustness with respect to the (c) systematic Rabi error and (d) the frequency drift error, where the solid-red and dashed-blue  lines denote the CSNHQC and DG  schemes, respectively.}
  \label{Fig7}
\end{figure}

To further illustrate the above statement, we also compare the performance of the $T$ gate, which can be constructed directly in our scheme, between our  CSNHQC scheme and the DG one. The dynamics construction of the T gate is presented in detail in the Appendix \ref{appendixB}. Figures \ref{Fig7}(a) and \ref{Fig7}(b) show that, for the $T$ gate, the optimal composite pulse sequence $N$ is 2, and the optimal detuning is $\Delta_{1}=17 \ \tilde{\Omega}_{1}$. We also plot the gate fidelities with respect to both the systematic Rabi error and the frequency drift error in Figs. \ref{Fig7}(c) and \ref{Fig7}(d), under the optimal value of $N$ and the detuning $\Delta_{1}$, respectively. Remarkably, our CSNHQC scheme is more robust than the DG one in the whole error range for both the systematic Rabi error and the frequency drift error.

\section{ PHYSICAL REALIZATION}

\begin{figure}[tbp]
  \centering
  \includegraphics[width=1\linewidth]{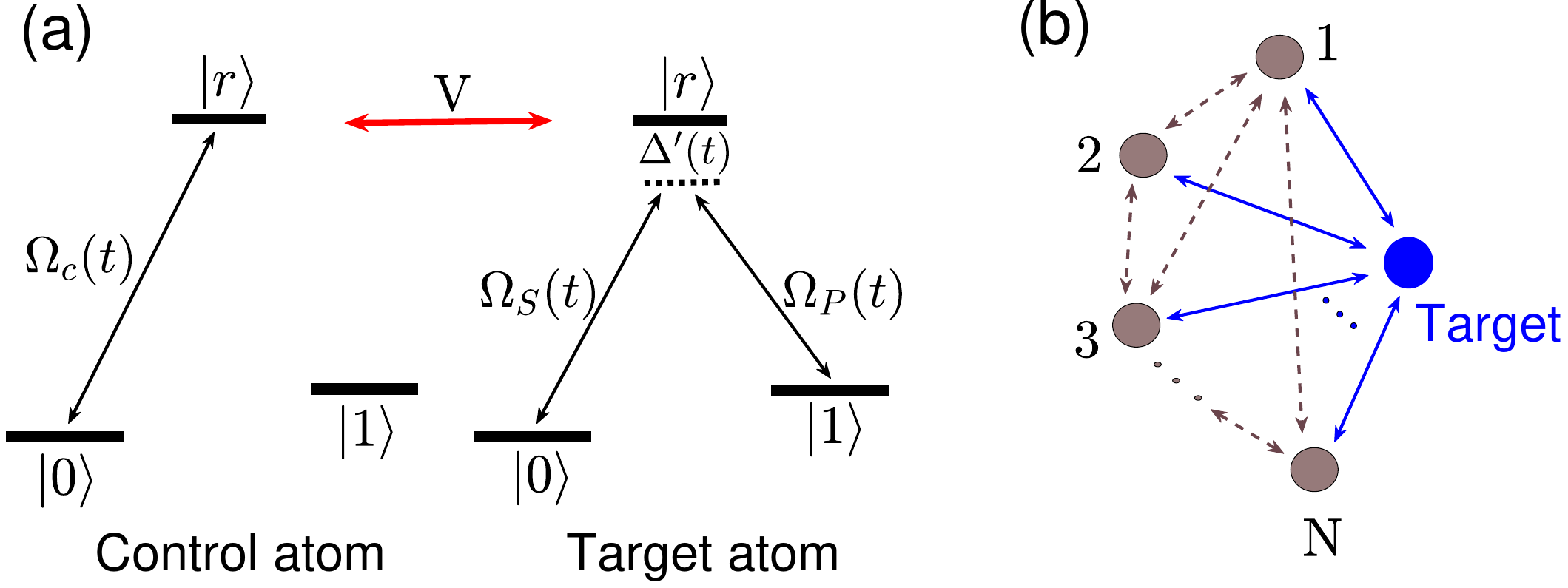}
  \caption{ (a) Illustration of the implementation of a two-qubit gate. The ground state $|0\rangle_c$ of the control atom is coupled resonantly to the Rydberg state $|r\rangle_c$ with Rabi frequencies $\Omega_c(t)$. For the target atom, ground states $|0\rangle_t$ and $|1\rangle_t$ are coupled to the Rydberg state $|r\rangle_t$ with Rabi frequencies $\Omega_S(t)$, $\Omega_P(t)$ and the same detuning $\Delta'(t)$, respectively. $V$ denotes the RRI strength. (b) Illustration of multiqubit controlled gate, which includes $N$ control atoms and one target atom.
  }\label{Fig8}
\end{figure}

Rydberg atoms with high principal quantum number possess excellent atomic properties including strong and long-range interaction, giant polarizability, and long lifetime \cite{ Saffman2010}. It provides a promising platform to take advantage of these properties to implement quantum gates between neutral atom qubits \cite{Jaksch2000, Isenhower2010, Levine2019,Zhaopeizi2017,Zhaopeizi2018,CPShen2019}. Our SNHQC scheme as well as the CSNHQC scheme can be readily implemented in the Rydberg atomic quantum system.  As for the single-qubit gate case, the required interaction in Eq. (\ref{7}) is just a conventional  three-level $\Lambda$ system driving by two external fields in the two-photon resonant way, which can be readily induced, see the interaction of the target atom in Fig. \ref{Fig8}(a). Thus, for the physical realization of our SNHQC scheme, in the following, we focus only on the construction of a two-qubit holonomic quantum gate.

The implementations of dynamical nontrivial two-qubit gate with two Rydberg atoms are based on the Rydberg blockade effect \cite{Jaksch2000}, where previous schemes can only be limited to the controlled-$\rm Z$ gate \cite{Jaksch2000} or controlled-not (CNOT) gate \cite{Isenhower2010, Levine2019}. Compared with previous dynamical schemes, our scheme can obtain arbitrary controlled-$U$ gate via choosing  different parameters. In addition, our scheme is more robust than the previous dynamical one in the presence of both the systematic Rabi error and the frequency drift error. Besides, even compared with the optimal situation of the dynamical approach, our scheme still has the advantage in terms of gate robustness. Moreover, our scheme can be extended easily to the case of multiqubit controlled gates, without increasing the gate time, which will benefit future large-scale quantum computation.

In Fig. \ref{Fig8}(a), we show the coupling configuration  for realizing the two-qubit quantum gate with Rydberg atoms. The qubit states
are represented by two long-lived hyperfine ground states $|0\rangle$ and $|1\rangle$, which can be manipulated by an optical Raman transition or a microwave field \cite{xia2015, Saffman2016}. The state $|0\rangle$ of the control atom is coherently coupled to the Rydberg state $|r\rangle$ by a focused laser field with Rabi frequency $\Omega_{\rm c}(t)$ in a resonant way. For the target atom, $|0\rangle$ ($|1\rangle$) is coupled to $|r\rangle$ with Rabi frequency $\Omega_S(t) [\Omega_P(t) ]$ and detuning $\Delta'(t)$. $V$ is the Rydberg-Rydberg interaction (RRI) strength depending on the interatomic distance and the principal quantum number $n$ of the involved Rydberg states. By these settings, the total Hamiltonian in two-qubit system is given by
 \begin{eqnarray}
\begin{array}{l}
\mathcal{H}_2(t)=\mathcal{H}_{\rm{c}}(t)+\mathcal{H}_{\rm{t}}(t)+\mathcal{H}_ V, \\
\mathcal{H}_{\rm{c}}(t)=[(1+\epsilon')\Omega_{\rm c}(t)|r\rangle_c\langle 0|+{\rm H.c.}]+\eta'|r\rangle_c\langle r|, \\
\mathcal{H}_{\rm{t}}(t)=[\Delta'(t)+\eta']|r\rangle_t\langle r|\\
\quad \quad \quad+(1+\epsilon')[\Omega_{S}(t)|r\rangle_t\langle 0|+\Omega_{P}(t)|r\rangle_t\langle 1|+\rm{H.c.}],\\
\mathcal{H}_V(t)=V|r\rangle_c\langle r|\otimes|r\rangle_t\langle r|,
\end{array}
\label{2qubit-total}
\end{eqnarray}
where  $\Omega_{\rm c}(t)=\bar{\Omega}_{\rm c}\cos(\omega t)$, $V=\omega$, and $\epsilon', \eta'$ denote the systematic Rabi error and the frequency drift error, respectively.

Here we suppose that the coupling strength of the control atom is much greater than that of the target atom, i.e., $\bar{\Omega}_{\rm c}\gg\Omega_m'$, and $V\gg \{\bar{\Omega}_{\rm c}, \Omega_m'\}$ with $\Omega_m'$ being the maximum value of $\sqrt{\Omega_{S}^2(t)+\Omega_{P}^2(t) }$ acting on the target atom. Note that, for large principal quantum number of Rydberg state and small interatomic distance, e.g., $n>100$ and $x<5\ \mu m$, the Rydberg-Rydberg interaction strength $V$ can be about $2\pi \times1000$ MHz \cite{Zhang2012,Petrosyan2017}. Therefore, an effective Hamiltonian can be obtained as \cite{wujinlei2021}
\begin{eqnarray}
\label{effective}
\mathcal{H}_{eff}(t)&=&|1\rangle_c\langle1|\otimes \mathcal{H}_{\rm{t}}(t).
\end{eqnarray}
It indicates that only when the control atom is in the state of $|1\rangle_c$, an interaction is applied on the target atom, and if the control atom is in the state of $|0\rangle_c$, the evolution of two-qubit system is frozen. By using this effective Hamiltonian we can achieve an arbitrary controlled  gate. Furthermore, as the Hamiltonian $\mathcal{H}_{\rm{t}}(t)$ of the target atom possesses the same form as in Eq. (\ref{7}), we can realize nontrivial two-qubit gates for our SNHQC scheme. The corresponding evolution operator in the two-qubit Hilbert space $\{|00\rangle, |01\rangle, |10\rangle, |11\rangle\}$ is
\begin{eqnarray}
\label{Utotal}
U(\tau')=
  \left(
\begin{array}{ccccccc}
 I             & 0           \\
 0             & U_1
\end{array}
\right),
\end{eqnarray}
where $\tau'$ is the total evolution time, and $U_1$ is given in Eq. (\ref{9}) with the basis being $|10\rangle$ and $|11\rangle$ now.

Next, we focus on the implementation of the CNOT gate and compare its performance with the conventional five-step DG scheme \cite{Isenhower2010}.  {Here we choose $|0\rangle\equiv|5S_{1/2}, F=1, m_F=0\rangle$ and $|1\rangle\equiv|5S_{1/2}, F=2,  m_F=0\rangle$ for the $^{87}$Rb atoms as two stable ground states, and $|r\rangle\equiv|83S,\;J=1/2, \; m_J=1/2\rangle$ as the Rydberg state \cite{MengLi2021}.
As $\gamma\neq\pi$, we have to use two controlled-$\sqrt{\rm X}$ gates to composite a CNOT gate. Nevertheless, considering the decoherence, our scheme still has more advantages than the traditional DG scheme. The decoherence operators are $\sigma^-_{i=c, t}=|0\rangle_i\langle r|+|1\rangle_i\langle r|$, $\sigma^z_{i=c, t}=|r\rangle_i\langle r|-|0\rangle_i\langle 0|-|1\rangle_i\langle 1|$ and $\sigma^{2}_{i=c, t}=|2\rangle_i\langle r|$, where $|2\rangle$ represents all the remaining Zeeman sublevels except for the computational states $|0\rangle$ and $|1\rangle$. For the sake of convenience, we suppose that decay rates of Rydberg state to the eight Zeeman-split ground levels are equal, and the corresponding decoherence rates are $\Gamma^-_{i=c, t}=\Gamma/8$, $\Gamma^z_{i=c, t}=\Gamma/10$ and $\Gamma^2_{i=c, t}=3\Gamma/4$, where $\Gamma=1/\tau$, and $\tau=200\ \mu$s is the lifetime of Rydberg state \cite{wujinlei2021}.}

\begin{figure}[t]
  \centering
  \includegraphics[width=0.95\linewidth]{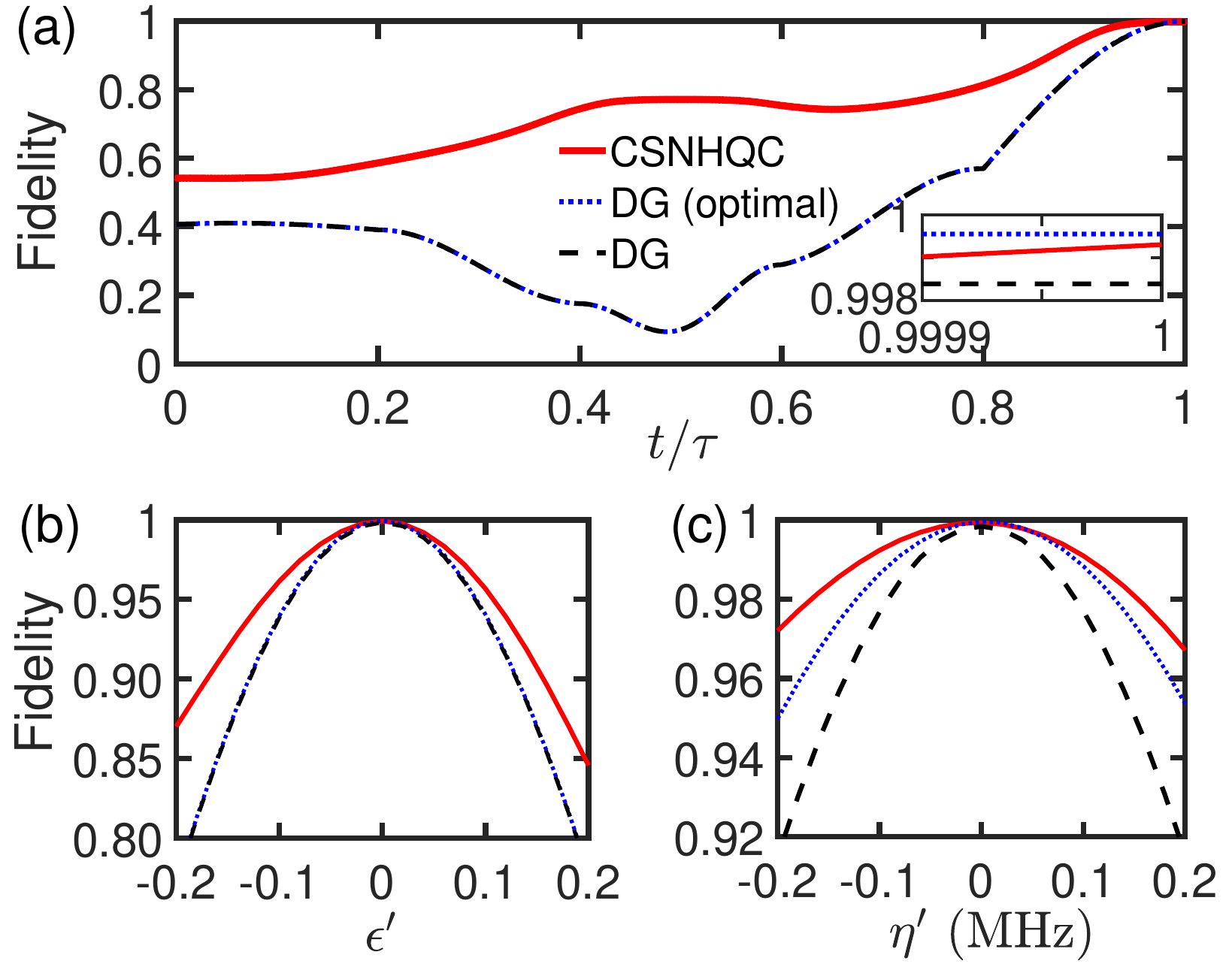}
  \caption{  Performance of  the implemented  CNOT gate. (a) The dynamics of the gate fidelities. The robustness of the CNOT gate in terms of the (b) systematic Rabi   and (c) the frequency drift errors. The solid red line, dashed black line and dotted blue line correspond to the CSNHQC scheme, DG scheme, and the optimized  DG scheme, respectively.}
 \label{Fig9}
\end{figure}

{To test our gate performance, we consider a general initial state $|\psi_2\rangle = [\cos\Theta_1 |0\rangle_c+\sin\Theta_1  {\rm exp}(i\Phi_1)|1\rangle_c]\otimes[\cos\Theta_2 |0\rangle_t+ \sin\Theta_2  {\rm exp}(i\Phi_2)|1\rangle_t]$, and define the two-qubit gate fidelity as $F_2= \langle\psi'_2(t)|\rho_2|\psi'_2(t)\rangle $ with the ideal final state being $|\psi'_2(t)\rangle$ and two-atom density operator being $\rho_2$. The parameters are set as $\bar{\Omega}_{\rm c}=2\pi\times 40$ MHz, $V=\omega=2\pi \times 500$ MHz and $\Omega_m'=2\pi$ MHz, which lead to a total operation time $\tau'=0.896 \ \rm{\mu s}$. When the average of the gate fidelity over 1296 input states is taken, as shown in Fig. \ref{Fig9}(a) (solid red line), the gate fidelity is $99.93\%$. We also test the fidelity of the dynamical CNOT gate, see Appendix \ref{appendixC} for details, with the same RRI strength and decoherence rates, but the Rabi frequency used in the DG scheme is a square pulse with the same value as the maximum value of our CSNHQC scheme. As shown in Fig. \ref{Fig9}(a), the fidelity of DG gate is $99.83\%$, which is smaller than our CSNHQC scheme. Considering that the DG scheme is not subject to the constraint of $\bar{\Omega}_{\rm c}\gg\Omega_m'$, we can increase the coupling strength of the target atom $\Omega_m'$ to optimize the DG scheme. Only when the coupling strength is 5 times the maximum value in our scheme, the fidelity of the DG scheme after optimization can be better than our scheme, and the gate fidelity is $99.95\%$.}

We now turn to examine the robustness of our CSNHQC CNOT gate. Figures \ref{Fig9}(b) and (c) demonstrate that our CSNHQC scheme (solid red line) are more robust than the DG scheme (dashed black line) in terms of both the systematic Rabi error and the frequency drift error. Even though compared with the optimal DG scheme (dotted blue line), our scheme is more advantageous in the whole error range.

In addition to the above nontrivial two-qubit gate, our scheme can readily be extended to the case of implementing  multiqubit gates. As shown in Fig. \ref{Fig8}(b), there are $N$ identical control atoms and one target atom. The level structure and transition-driven fields of control atoms and target atom are the same as the two-qubit gate case as shown in Fig. \ref{Fig8}(a). When the condition $V\gg\bar{\Omega}_{\rm c}\gg \Omega_m'$ is satisfied, the effective Hamiltonian of the multiqubit system is \cite{wujinlei2021}
\begin{eqnarray}
\label{multi-effective}
\mathcal{H}'_{eff}(t)&=&\left(\otimes_j^N|1\rangle_j\langle1|\right )\otimes \mathcal{H}_{\rm{t}}(t).
\end{eqnarray}
Therefore, we can realize the shortest path $N$-qubit controlled gate without increasing the operation time.

\section{DISCUSSION AND CONCLUSION}
In summary, we propose to realize the SNHQC via the inverse engineering of the Hamiltonian. The single-qubit gate fidelities of $\rm S$ gate, $\rm T$ gate, and $\sqrt{\rm H}$ gate can be as high as $99.97\%$, $99.99\%$, and $99.97\%$ under the decoherence effect. Besides, these gates are more robust than the traditional NHQC scheme against both the systematic Rabi error and frequency drift error. 
Moreover, the gate performance  can be further  improved by the proposed composite dynamical decoupling pulse technique. In the absence of decay between $|0\rangle$ and $|1\rangle$, as the composite pulse sequences number \emph{N} increases, the population of the axillary excited state decrease towards zero, and the robustness with respect to the frequency drift error becomes better and better. This improvement means our scheme can exceed the optimal dynamical scheme in  certain decoherence rate ranges.

Furthermore, we construct the nontrivial two-qubit gate in two Rydberg atoms via Rydberg blockade and inverse engineering of the Hamiltonian. For the CNOT gate, the robustness of our scheme against the systematic Rabi error and frequency drift error are much
stronger than that of the traditional dynamical scheme. Even though compared with the optimized dynamical scheme, our scheme is still more robust in terms of both errors. In addition, our scheme can also be extended to the construction of multiqubit controlled gates without increasing the operation time, providing an alternative strategy for future scalable quantum computing. Finally, as the required interaction in Eq. (\ref{7}) is the conventional  three-level $\Lambda$ system driving by two external fields in a two-photon resonant way, our scheme can be extended directly to other systems, e.g., superconducting circuit, cavity QED, quantum dots, trapped ions, etc.

\begin{acknowledgements}
This work was supported by the Key-Area Research and Development Program of GuangDong Province (Grant No. 2018B030326001), the National Natural Science Foundation of China (Grant No. 11874156), and  Science and Technology Program of Guangzhou (Grant No. 2019050001).

\end{acknowledgements}

\begin{figure}[tb]
  \centering
  \includegraphics[width=0.95\linewidth]{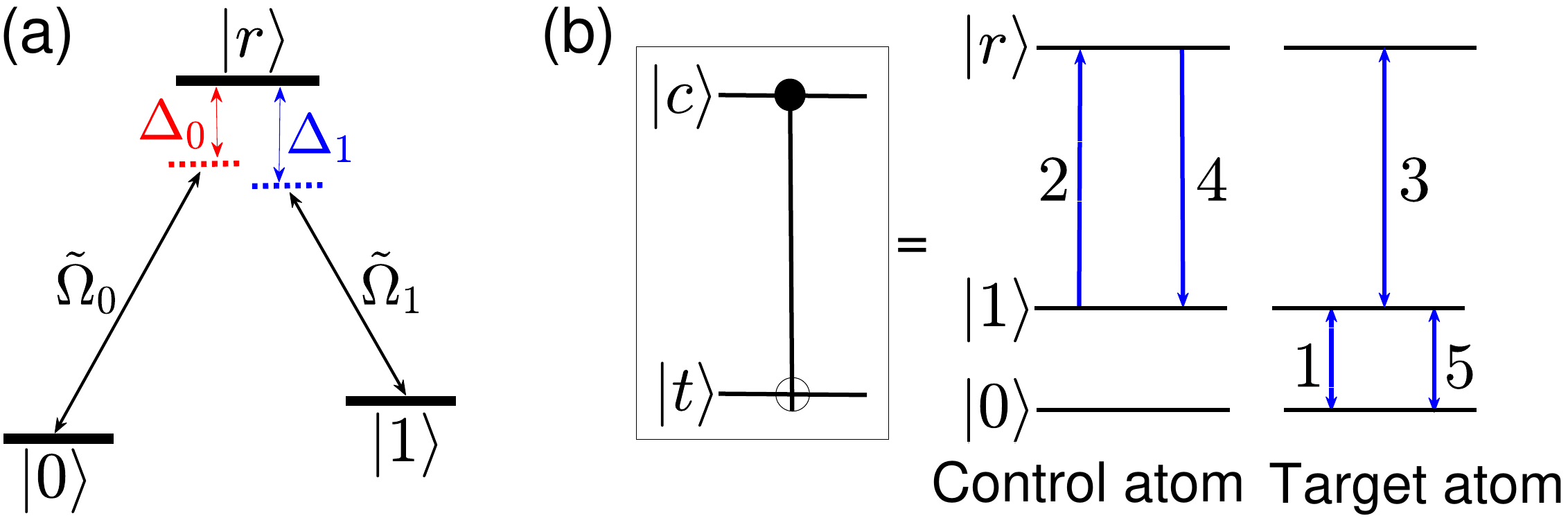}
\caption{Illustration of the construction of dynamical quantum gates, with $|0\rangle$ and $|1\rangle$ being two ground states to encode quantum information, while $|r\rangle$ denotes an auxiliary Rydberg state. (a) Coupling configuration for $H$ and $T$ gates. The ground states $|0\rangle$ and $|1\rangle$ are coupled to the excited state $|r\rangle $ with Rabi frequencies $\tilde{\Omega}_0$ and $\tilde{\Omega}_1$, and with detuning $\Delta_0$ and $\Delta_1$, respectively. (b)  Implementation of the two-qubit CNOT gate based on Rydberg blockade with five pulse sequences.   }\label{Fig10}
\end{figure}

\appendix
\section{THE CONSTRUCTION OF THE DYNAMICAL HADAMARD GATE} \label{appendixA}

The coupling configuration for the dynamical H gate is shown in Fig. \ref{Fig10}(a). The three-level atom consists of two ground states $|0\rangle$, $|1\rangle$ and a Rydberg excited state $|r\rangle$. The transitions of $|0\rangle\rightarrow|r\rangle$ and $|1\rangle\rightarrow|r\rangle$ are driven by two lasers with Rabi frequencies $\tilde{\Omega}_0$ and $\tilde{\Omega}_1$, and with detuning $\Delta_0$ and $\Delta_1$, respectively. In the interaction picture with respective to the free energy of the atom, the interaction Hamiltonian is
\begin{eqnarray}
\label{dynamicalH}
\tilde{H}(t)&=&(1+\epsilon)(\tilde{\Omega}_0e^{i\varphi'_0}|r\rangle\langle0|e^{-i\Delta_{0}t} \notag \\
&+&\tilde{\Omega}_1e^{i\varphi'_1}|r\rangle\langle1|e^{-i\Delta_{1}t}+ \text{H.c.})
+\eta\tilde{\Omega}_0|r\rangle\langle r|,
\end{eqnarray}
where $\tilde{\Omega}_0=\left[\int^{\tau}_0\Omega_0(t)dt+\int^{\tau}_0\Omega_1(t)dt\right]/(2\tau)$ equals to the average of Rabi frequency of the CS-NHQC scheme, $\epsilon$ and $\eta$ represent the systematic Rabi error fraction and the frequency drift error fraction, respectively. The effective Hamiltonian in the case of large detuning is
\begin{eqnarray}
\label{dynamical-effective}
\tilde{H}_{eff}(t)&=& \frac{\tilde{\Omega}^2_0}{\Delta_{0}}(|0\rangle\langle0|-|r\rangle\langle r|) +\frac{\tilde{\Omega}^2_1}{\Delta_{1}}(|1\rangle\langle1|-|r\rangle\langle r|) \notag \\
&+&\frac{\tilde{\Omega}_0\tilde{\Omega}_1}{2\Delta_{01}}(|0\rangle\langle1|e^{i(\Delta_{0}-\Delta_{1})t} +\rm{H.c.}),
\end{eqnarray}
where we have set $1/\Delta_{01}=  1/\Delta_{0}+1/\Delta_{1}$, $\epsilon=\eta=0$ and $\varphi'_0=\varphi'_1=0$. When we choose $\tilde{\Omega}^2_0/\Delta_{0}=\tilde{\Omega}^2_1/\Delta_{1}$ and in the computation space of Span$\{|0\rangle, |1\rangle\}$, the effective Hamiltonian reduces to
\begin{eqnarray}
\label{dynamical-effectiver}
\tilde{H}_{eff}(t)&=&\frac{\tilde{\Omega}_0\tilde{\Omega}_1}{2\Delta_{01}} (|0\rangle\langle1|e^{i(\Delta_{0}-\Delta_{1})t} +\rm{H.c.}).
\end{eqnarray}
Then in the frame of $\exp{(iht)}$ with $h=(\Delta_{0}-\Delta_{1})(|0\rangle\langle0|-|1\rangle\langle1|)/2$, Eq. (\ref{dynamical-effectiver}) becomes
\begin{eqnarray}
\label{dynamical-effectiveH}
\tilde{H}'_{eff}(t)&=&\frac{\tilde{\Omega}_0\tilde{\Omega}_1}{2\Delta_{01}}\bm{\sigma}_x+ \frac{\Delta_{0}-\Delta_{1}}{2}\bm{\sigma}_z,
\end{eqnarray}
where $\bm{\sigma}_{i=x,z}$ are the Pauli operators composed of two basis $|0\rangle$ and $|1\rangle$. Furthermore, we choose $(\tilde{\Omega}_0\tilde{\Omega}_1)/\Delta_{01}= \Delta_{0}-\Delta_{1} = \sqrt{2}\Omega_{eff}$,  the corresponding evolution operator is
\begin{eqnarray}
\label{dynamical-effectiveH}
U&=&\cos\theta'\rm {\textbf{I}}-\textit{i}\sin\theta' \rm{\textbf{H}},
\end{eqnarray}
with $\theta'=\Omega_{eff}t$. Therefore, we can implement the H gate in one step by setting $\theta'= \pi/2$. To optimize the dynamical H gate, we set $\Delta_{1}=j\tilde{\Omega}_1 (j\geq0)$, and find the optimal value of $j$ that maximizes the fidelity of the dynamical H gate under decoherence, as shown in Fig. \ref{Fig6}(b) in the maintext.

\section{THE CONSTRUCTION OF THE DYNAMICAL $T$ GATE}  \label{appendixB}
To achieve the dynamical T gate in a three-level $\Lambda$ system, we also adopt the system as shown in Fig. \ref{Fig10}(a). Different from the H gate, we set $\Delta_{0}=\Delta_{1}$ and $\tilde{\Omega}_0=\tilde{\Omega}_1$ with $ \tilde{\Omega}_1=\int^{\tau}_0\Omega_1(t)dt/\tau$ being the average of Rabi frequency of the CS-NHQC scheme. Therefore, in this case, the effective Hamiltonian in Eq. (\ref{dynamical-effectiver}) reduces to
\begin{eqnarray}
\label{dynamical-effectiveT}
\tilde{H}_{eff}''(t)&=&\frac{\tilde{\Omega}_1^2}{\Delta_1}(|0\rangle\langle1|e^{i\varphi'} +|1\rangle\langle0|e^{-i\varphi'}).
\end{eqnarray}
with $\varphi'=\varphi'_1-\varphi'_0$. The construction of the T gate is completed in three steps, i.e., T$=U_3 U_2 U_1$, and the three steps correspond to (1) $\varphi'=0$ and $\tilde{\Omega}_1^2\tau_1/\Delta_1=\pi/4$, (2) $\varphi'=3\pi/2$ and $ \tilde{\Omega}_1^2\tau_2/\Delta_1 =\pi/8$, (3) $\varphi'=\pi$ and $ \tilde{\Omega}_1^2\tau_3/\Delta_1 =\pi/4$, respectively.
For the goal of optimizing the dynamical gate scheme, we set $\Delta_1=k\tilde{\Omega}_1 (k\geq0)$, and find the value of $k$ that maximizes the fidelity of the T gate in the presence of decoherence, as shown  in Fig. \ref{Fig7}(b) in the maintext.

\begin{figure}[thb]
  \centering
  \includegraphics[width=0.85\linewidth]{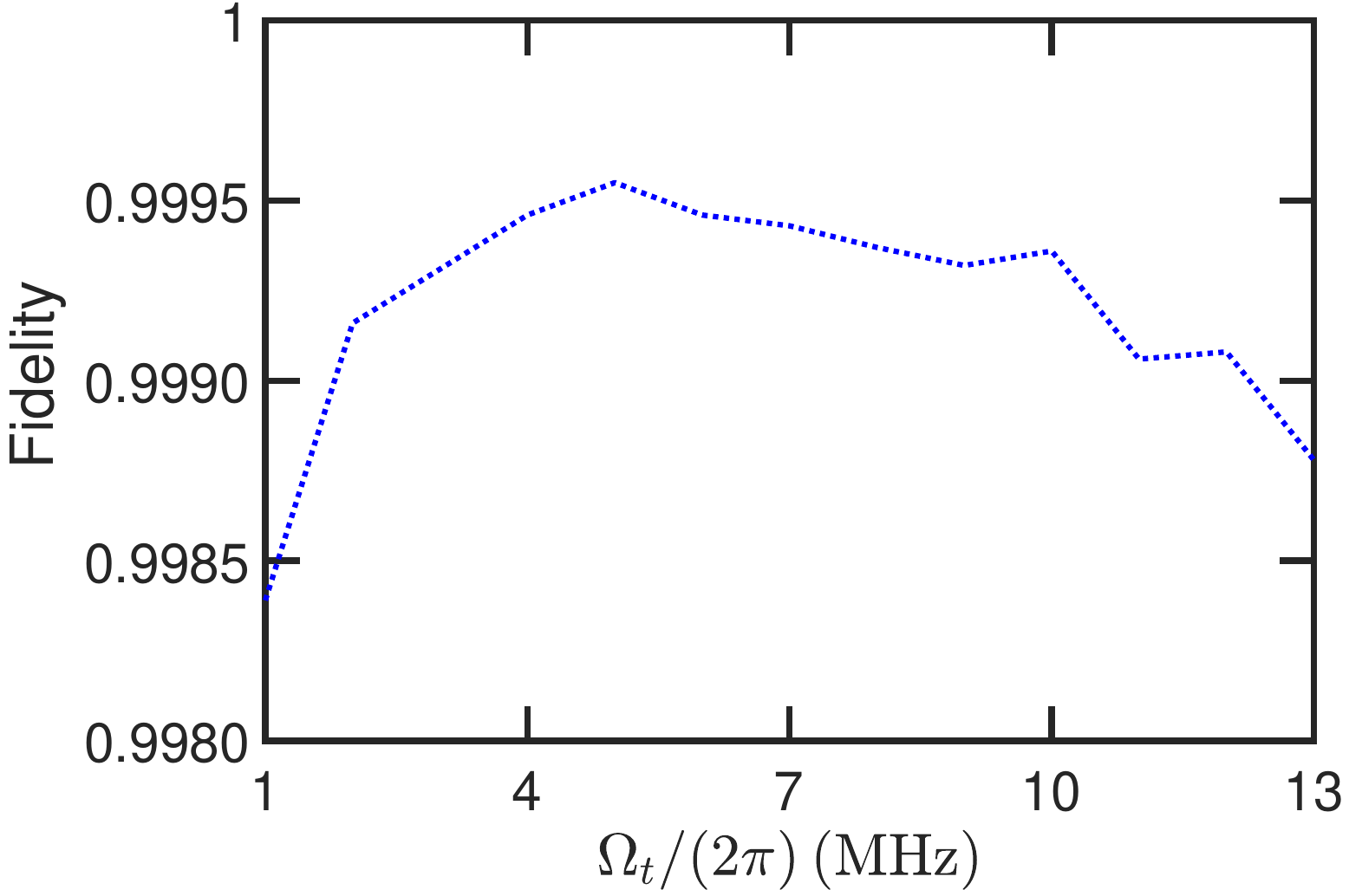}
\caption{ The fidelity of CNOT gate with respect to Rabi frequency $\Omega_t$ of the target atom in step 3. Other parameters are $\Omega_{01}=\pi$ MHz, $\Omega_c=2\pi\times 40 $ MHz and $V=2\pi\times 500 $ MHz. The decoherence rates are $\Gamma^-=\Gamma/8 $, $\Gamma^z=\Gamma/10$ and $\Gamma^2=3\Gamma/4$ with $\Gamma=1/\tau$ and $\tau=200 \mu$s. }
\label{Fig11}
\end{figure}

\section{THE CONSTRUCTION OF THE DYNAMICAL CNOT GATE}  \label{appendixC}

The standard approach for constructing CNOT gates for two Rydberg atomic qubits using Rydberg blockade is to perform Hadamard rotations on the target qubit before and after the two-qubit controlled phase gate \cite{Isenhower2010}. The CNOT gate procedure is carried out in five steps as shown in Fig. \ref{Fig10}(b). The Hadamard rotations on two qubit-states $|0\rangle$ and $|1\rangle$  can be obtained by
an optical Raman transition or a microwave field, and are applied on the target atom in steps 1 and 5, the corresponding Hamiltonian is written as
\begin{eqnarray}
\label{dyn_H15}
H_{1,5}&=&[(1+\epsilon')\Omega_{01}|1\rangle_t\langle 0|+\rm{H.c.}]+\eta'|1\rangle_t\langle 1|,
\end{eqnarray}
where $\Omega_{01}=\pi$ MHz is the Rabi frequency for the target qubit-state transition $|0\rangle_t\leftrightarrow|1\rangle_t$; $\epsilon'\in[-0.2, 0.2]$ and $\eta'\in[-0.2 , 0.2]$ MHz are the systematic Rabi error fraction and the strength of frequency drift error, respectively.

In step 2, the resonant pulse $\Omega_{c}$ is applied on control atom, which drives the transition of $|1\rangle_c\leftrightarrow|r\rangle_c$, and the corresponding Hamiltonian reads
\begin{eqnarray}
\label{dyn_H2}
H_{2}&=&[(1+\epsilon')\Omega_{c}|r\rangle_c\langle 1|+\rm{H.c.}]+\eta'|r\rangle_c\langle r|.
\end{eqnarray}
We here choose $\Omega_{c}=2\pi\times 40$ MHz, which is the maximum value of the used pulse in the CS-NHQC scheme.

We turn off the resonant pulse on the control atom and apply another pulse with Rabi frequency $\Omega_{t}$ on the target atom in step 3. The Hamiltonian is
\begin{eqnarray}
\label{dyn_H3}
H_{3}&=&[(1+\epsilon')\Omega_{t}|r\rangle_t\langle 1|+\rm{H.c.}]+\eta'|r\rangle_t\langle r|.
\end{eqnarray}
When it satisfies $\Omega_{t}t=\pi$, the transition $|1\rangle_t\rightarrow|r\rangle_t\rightarrow-|1\rangle_t$ can be realized. However, if the control atom is in $|r\rangle_c$ state, this transition will be blocked due to the existence of the Rydberg-Rydberg interaction. The  interaction Hamiltonian reads
\begin{eqnarray}
\label{dyn_H3}
H_{V}&=&V|r\rangle_c\langle r|\otimes|r\rangle_t\langle r|,
\end{eqnarray}
where $V\gg\Omega_{t}$.

Subsequently, we turn off the resonant pulse on the target atom and perform another pulse with Rabi frequencies $\Omega_{c}$ on the control atom in step 4 to achieve the transition $|r\rangle_c\rightarrow|1\rangle_t$, where the Hamiltonian is the same structure as step 2.

Next, we will discuss the third step in more detail. In order to compare the performance of the DG scheme and the CS-NHQC scheme under the same conditions, we take the square pulse value of the DG scheme as $\Omega_{t}= 2\pi$ MHz, which is the maximum value of the CS-NHQC scheme, and the corresponding numerical results are shown in Fig. \ref{Fig9} in the maintext. It shows that under the same maximum Rabi frequency, the DG scheme is far inferior to the CS-NHQC scheme. However, the DG scheme is not restricted by the condition that $\Omega_{c}$ is much greater than $\Omega_{t}$, which means that $\Omega_{t}$ in the DG scheme can be larger to improve the gate fidelity. But, the DG scheme is still restricted by the condition of $V\gg\Omega_{t}$. Therefore, we set $V= 2\pi \times 500$ MHz, the parameters in other steps remain unchanged, and then find the value of $\Omega_{t}$ that can maximize the fidelity of the DG scheme. Figure \ref{Fig11} shows that when $\Omega_{t}= 2\pi\times 5$ MHz, the DG scheme achieves the optimal result. Nevertheless, our CS-NHQC scheme is still more robust than the optimal DG scheme, as shown in Figs.  \ref{Fig9}(b) and  \ref{Fig9}(c) in the maintext.


\begin{thebibliography}{99}

\bibitem{Nielson}
M. A. Nielsen and I. L. Chuang,
\emph{Quantum computation and quantum information} (Cambridge University Press, Cambridge, 2000).


\bibitem{cavity}
Q. A. Turchette, C. J. Hood, W. Lange, H. Mabuchi, and H. J. Kimble,
Measurement of conditional phase shifts for quantum logic,
Phys. Rev. Lett. \textbf{75}, 4710 (1995).

\bibitem{ions}
J. I. Cirac and P. Zoller,
Quantum computations with cold trapped ions,
Phys. Rev. Lett. \textbf{74}, 4091 (1995).

\bibitem{atom1}
G. K. Brennen, C. M. Caves, P. S. Jessen, and I. H. Deutsch,
Quantum logic gates in optical lattices,
Phys. Rev. Lett. \textbf{82}, 1060 (1999).

\bibitem{atom2}
D. Jaksch, H. J. Briegel, J. I. Cirac, C. W. Gardiner, and P. Zoller,
Entanglement of atoms via cold controlled collisions,
Phys. Rev. Lett. \textbf{82}, 1975 (1999).

\bibitem{Saffman2010}
M. Saffman, T. G. Walker, and K. M{\o}lmer,
Quantum information with Rydberg atoms,
Rev. Mod. Phys. \textbf{82}, 2313 (2010).

\bibitem{hqc} P. Zanardi and M. Rasetti,
Holonomic quantum computation,
Phys. Lett. A {\bf 264}, 94 (1999).


\bibitem{Solinas2004}
P. Solinas, P. Zanardi, and N. Zangh\`{\i},
Robustness of non-Abelian holonomic quantum gates against parametric noise,
Phys. Rev. A \textbf{70}, 042316 (2004).


\bibitem{Solinas2012}
P. Solinas, M. Sassetti, T. Truini, and N. Zangh\`{\i},
On the stability of quantum holonomic gates,
New J. Phys. \textbf{14}, 093006 (2012).


\bibitem{Johansson2012}
M. Johansson, E. Sj\"{o}qvist, L. M. Andersson, M. Ericsson, B. Hessmo, K. Singh, and D.-M. Tong,
Robustness of nonadiabatic holonomic gates,
 Phys. Rev. A \textbf{86}, 062322 (2012).

\bibitem{JP99} J. Pachos, P. Zanardi, and M. Rasetti,
Non-Abelian Berry connections for quantum computation,
Phys. Rev. A {\bf 61}, 010305(R) (1999).


\bibitem{Duan2001}
L.-M. Duan, J. I. Cirac, and P. Zoller,
 Geometric manipulation of trapped ions for quantum computation,
  Science \textbf{292}, 1695 (2001).

%
%

\bibitem{Sjoqvist2012}
E. Sj\"{o}qvist, D.-M. Tong, L. M. Andersson, B. Hessmo, M. Johansson, and K. Singh,
 Non-adiabatic holonomic quantum computation,
 New J. Phys \textbf{14}, 103035 (2012).

 \bibitem{xu2012}
G.-F. Xu, J. Zhang, D.-M. Tong, E. Sj\"{o}qvist, and L. C. Kwek,
Nonadiabatic holonomic quantum computation in decoherence-free subspaces,
 Phys. Rev. Lett. \textbf{109}, 170501 (2012).

\bibitem{Abdumalikov2013}
A. A. Abdumalikov, J. M. Fink, K. Juliusson, M. Pechal, S. Berger, A. Wallraff, and S. Filipp,
Experimental realization of non-Abelian non-adiabatic geometric gates,
Nature (London) \textbf{496}, 482 (2013).

\bibitem{long2013}
G. Feng, G. Xu, and G. Long,
Experimental Realization of Nonadiabatic Holonomic Quantum Computation,
Phys. Rev. Lett. {\bf 110}, 190501 (2013).

\bibitem{Duan2014}
C. Zu, W.-B. Wang, L. He, W.-G. Zhang, C.-Y. Dai, F. Wang, and L.-M. Duan,
Experimental realization of universal geometric quantum gates with solid-state spins,
Nature (London) {\bf 514}, 72 (2014).

\bibitem{SAC2014}
S. Arroyo-Camejo, A. Lazariev, S. W. Hell, and G. Balasubramanian,
Room temperature high-fidelity holonomic single-qubit gate on a solid-state spin,
Nat. Commun. {\bf 5}, 4870 (2014).

\bibitem{SDanilin2018}
S. Danilin, A. Veps\"{a}l\"{a}inen, and G. S. Paraoanu,
Experimental state control by fast nonAbelian holonomic gates with a
superconducting qutrit,
Phys. Scr. \textbf{93}, 055101 (2018).

\bibitem{xu2015}
G.-F. Xu, C.-L. Liu, P.-Z. Zhao, and D.-M. Tong,
Nonadiabatic holonomic gates realized by a single-shot implementation,
Phys. Rev. A \textbf{92}, 052302 (2015).

\bibitem{ESjoqvist2016}
E. Sj\"{o}qvist,
Nonadiabatic holonomic single-qubit gates in off-resonant systems,
Phys. Lett. A \textbf{380}, 65 (2016).




\bibitem{Herterich2016}
E. Herterich and E. Sj\"{o}qvist,
Single-loop multiple-pulse nonadiabatic holonomic quantum gates,
 Phys. Rev. A \textbf{94}, 052310 (2016).

 \bibitem{hong2018}
Z.-P. Hong, B.-J. Liu, J.-Q. Cai, X.-D. Zhang, Y. Hu, Z.-D. Wang, and Z.-Y. Xue,
 Implementing universal nonadiabatic holonomic quantum gates with transmons,
 Phys. Rev. A \textbf{97}, 022332 (2018).

\bibitem{Sekiguchi2017}
Y. Sekiguchi, N. Niikura, R. Kuroiwa, H. Kano, and H. Kosaka,
 Optical holonomic single quantum gates with a geometric spin under a zero field,
 Nat. Photonics \textbf{11}, 309 (2017).

\bibitem{long2017}
 H. Li, L. Yang, and G. Long,
Experimental realization of single-shot nonadiabatic holonomic gates in nuclear spins,
Sci. China: Phys., Mech. Astron. {\bf 60}, 080311 (2017).


\bibitem{zhou2017}
B.-B. Zhou, P. C. Jerger, V. O. Shkolnikov, F. J. Heremans, G. Burkard, and D. D. Awschalom,
Holonomic Quantum Control by Coherent Optical Excitation in Diamond,
Phys. Rev. Lett. {\bf 119}, 140503 (2017).

\bibitem{sun2018}
 Y. Xu, W. Cai, Y. Ma, X. Mu, L. Hu, T. Chen, H. Wang, Y.-P. Song, Z.-Y. Xue, Z.-Q. Yin, and L. Sun,
Single-Loop Realization of Arbitrary Nonadiabatic Holonomic Single-Qubit Quantum Gates in a Superconducting Circuit,
Phys. Rev. Lett. {\bf 121}, 110501 (2018).

\bibitem{NI2018}
 N. Ishida, T. Nakamura, T. Tanaka, S. Mishima, H. Kano, R. Kuroiwa, Y. Sekiguchi, and H. Kosaka,
Universal holonomic single quantum gates over a geometric spin with phase-modulated polarized light,
Opt. Lett. {\bf 43}, 2380 (2018).

\bibitem{peng2019} Z. Zhu, T. Chen, X. Yang, J. Bian, Z.-Y. Xue, and X. Peng,
Single-Loop and Composite-Loop Realization of Nonadiabatic Holonomic Quantum Gates in a Decoherence-Free Subspace,
Phys. Rev. Appl. {\bf 12}, 024024 (2019).


\bibitem{Carlini2012}
A. Carlini and T. Koike,
Time-optimal transfer of coherence,
Phys. Rev. A \textbf{86}, 054302 (2012).

\bibitem{Carlini2013}
A. Carlini and T. Koike,
 Time-optimal unitary operations in ising chains: unequal couplings and fixed fidelity,
J. Phys. A \textbf{46}, 045307 (2013).

\bibitem{wamg2015}
X.-T. Wang, M. Allegra, and K. Jacobs,
Quantum brachistochrone curves as geodesics: Obtaining accurate minimum-time protocols for the control of quantum systems,
Phys. Rev. Lett. \textbf{114}, 170501 (2015).

\bibitem{geng2016}
J. Geng, Y. Wu, and X.-T. Wang,
 Experimental time-optimal universal control of spin qubits in solids,
  Phys. Rev. Lett. \textbf{117}, 170501 (2016).

\bibitem{liuarxiv}
B.-J. Liu, Z.-Y. Xue, and M.-H. Yung,
Briachistochronic non-adiabatic holonomic quantum control,
arXiv:2001.05182.

\bibitem{chen2020}
T. Chen, P. Shen, and Z.-Y. Xue,
Robust and fast holonomic quantum gates with encoding on superconducting circuits,
Phys. Rev. Appl.\textbf{ 14}, 034038 (2020).


\bibitem{jiln2021} L.-N. Ji, C.-Y. Ding, T. Chen, and Z.-Y. Xue,
Noncyclic Geometric Quantum Gates with Smooth Paths via Invariant-Based Shortcuts,
Adv. Quantum Technol. {\bf 4}, 2100019 (2021).

\bibitem{shenp2021} P. Shen, T. Chen, and Z.-Y. Xue,
Ultrafast holonomic quantum gates,
Phys. Rev. Appl. {\bf 16}, 044004 (2021).


\bibitem{yu2020} Z. Han, Y. Dong, B. Liu, X. Yang, S. Song, L. Qiu, D. Li, J. Chu, W. Zheng, J. Xu, \emph{et al}.,
Experimental Realization of Universal Time-optimal non-Abelian Geometric Gates,
arXiv:2004.10364 (2020).

\bibitem{sunfw2021} Y. Dong, C. Feng, Y. Zheng, X.-D. Chen, G.-C. Guo, and F.-W. Sun,
Fast high-fidelity geometric quantum control with quantum brachistochrones,
Phys. Rev. Res. {\bf 3}, 043177 (2021).


\bibitem{zhu2003} S.-L. Zhu and Z.-D. Wang,
Unconventional Geometric Quantum Computation,
Phys. Rev. Lett. {\bf 91}, 187902 (2003).

\bibitem{du2006} J. Du, P. Zou, and Z.-D. Wang,
Experimental implementation of high-fidelity unconventional geometric quantum gates using an NMR interferometer,
Phys. Rev. A  {\bf 74}, 020302(R) (2006).


\bibitem{xu2018}
G.-F. Xu, D.-M. Tong, and E. Sj\"{o}qvist,
Path-shortening realizations of nonadiabatic holonomic gates,
 Phys Rev A \textbf{98}, 052315 (2018).

\bibitem{zhao2020}
P.-Z. Zhao, K.-Z. Li, G.-F. Xu, and D.-M. Tong,
General approach for constructing Hamiltonians for nonadiabatic holonomic quantum computation,
 Phys. Rev. A \textbf{101}, 062306 (2020).

 \bibitem{SaiLi2020}
 S. Li, P. Shen, T. Chen, and Z.-Y. Xue,
Noncyclic nonadiabatic holonomic quantum gates via shortcuts to adiabaticity,
Front. Phys. \textbf{16}, 51502 (2021).


\bibitem{xugf2014}  G. Xu and G. Long,
Protecting geometric gates by dynamical decoupling,
Phys. Rev. A {\bf 90}, 022323 (2014).


\bibitem{liubj2019} B.-J. Liu, X.-K. Song, Z.-Y. Xue, X. Wang, and M.-H.Yung,
Plug-and-Play Approach to Nonadiabatic Geometric Quantum Gates,
Phys. Rev. Lett. {\bf 123}, 100501 (2019).


\bibitem{Lisai2020}
S. Li, T. Chen, and Z.-Y. Xue,
Fast holonomic quantum computation on superconducting circuits with optimal control,
Adv. Quantum Technol. \textbf{3}, 2000001 (2020).

\bibitem{liubj2021} B.-J. Liu, Y.-S. Wang, and M.-H. Yung,
 Super-robust nonadiabatic geometric quantum control,
  Phys. Rev. Res. {\bf 3}, L032066 (2021).

\bibitem{Lisai2021} S. Li and Z.-Y. Xue,
Dynamically corrected nonadiabatic holonomic quantum gates,
Phys. Rev. Appl. {\bf 16}, 044005 (2021).


\bibitem{yudp2019} T. Yan, B.-J. Liu, K. Xu, C. Song, S. Liu, Z. Zhang,
H. Deng, Z. Yan, H. Rong, K. Huang, M.-H. Yung, Y.
Chen, and D. Yu, Experimental Realization of Nonadiabatic
Shortcut to Non-Abelian Geometric Gates,
Phys. Rev. Lett. {\bf 122}, 080501 (2019).


\bibitem{YS2019} Y. Sekiguchi, Y. Komura, and H. Kosaka,
Dynamical Decoupling of a Geometric Qubit,
Phys. Rev. Appl. {\bf 12}, 051001 (2019).

\bibitem{aimz2020}
M.-Z. Ai, S. Li, Z. Hou, R. He, Z.-H. Qian, Z.-Y. Xue, J.-M. Cui, Y.-F. Huang, C.-F. Li, and G.-C. Guo,
Experimental realization of nonadiabatic holonomic single-qubit quantum gates with optimal control in a trapped ion,
Phys. Rev. Appl. \textbf{14}, 054062 (2020).

\bibitem{aimz2021}
M.-Z. Ai, S. Li, R. He, Z.-Y. Xue, J.-M. Cui, Y.-F. Huang, C.-F. Li, and G.-C. Guo,
Experimental realization of nonadiabatic holonomic single-qubit quantum gates with two dark paths in a trapped ion,
Fundamental Research (2021).


\bibitem{sunfwprappl2021} Y. Dong, S.-C. Zhang, Y. Zheng, H.-B. Lin, L.-K. Shan, X.-D. Chen, W. Zhu, G.-Z. Wang, G.-C. Guo, and F.-W. Sun,
Experimental implementation of universal holonomic quantum computation on solid-state spins with optimal control,
Phys. Rev. Appl. {\bf 16}, 024060 (2021).

\bibitem{xuy2021}  S. Li, B.-J. Liu, Z. Ni,  L. Zhang, Z.-Y. Xue, J. Li, F. Yan, Y. Chen, S. Liu, M.-H. Yung, Y. Xu, and D. Yu,
Realization of Super-Robust Geometric Control in a Superconducting Circuit,
Phys. Rev. Appl. {\bf 16}, 064003 (2021).



\bibitem{Kang2018}
 Y.-H. Kang, Y.-H. Chen, Q.-C. Wu, B.-H. Huang, Y. Xia, and J. Song,
Reverse engineering of a Hamiltonian by designing the evolution operators,
Sci. Rep. \textbf{6}, 30151 (2016).



\bibitem{Odelin2019}
D. Gu\'{e}ry-Odelin, A. Ruschhaupt, A. Kiely, E. Torrontegui, S. Mart\'{\i}nez-Garaot, and J. G. Muga,
Shortcuts to adiabaticity: Concepts, methods, and applications,
Rev. Mod. Phys. \textbf{91}, 045001 (2019).

\bibitem{PHLeung2018}
P. H. Leung, K. A. Landsman, C. Figgatt, N. M. Linke, C. Monroe, and K. R. Brown,
Robust 2-Qubit Gates in a Linear Ion Crystal Using a Frequency-Modulated Driving Force,
Phys. Rev. Lett. {\bf 120}, 020501 (2018).

\bibitem{CFiggatt2019}
C. Figgatt, A. Ostrander, N. M. Linke, K. A. Landsman, D. Zhu, D. Maslov, and C. Monroe,
Parallel entangling operations on a universal ion-trap quantum computer,
Nature \textbf{572}, 368$-$372 (2019).

\bibitem{LandsmanKA2019}
K. A. Landsman, Y. Wu, P. H. Leung, D. Zhu, N. M. Linke, K. R. Brown, L. Duan, and C. Monroe,
Two-qubit entangling gates within arbitrarily long chains of trapped ions,
Phys. Rev. A \textbf{100}, 022332 (2019).






\bibitem{Viola1999}
L. Viola, E. Knill, and S. Lloyd,
Dynamical decoupling of open quantum systems,
Phys. Rev. Lett. \textbf{82}, 2417 (1999).





\bibitem{Jaksch2000}
D. Jaksch, J. I. Cirac, P. Zoller, S. L. Rolston, R. C\^{o}t\'{e}, and M. D. Lukin,
Fast quantum gates for neutral atoms,
Phys. Rev. Lett. \textbf{85}, 2208 (2000).

%

\bibitem{Isenhower2010}
L. Isenhower, E. Urban, X.-L. Zhang, A. T. Gill, T. Henage, T. A. Johnson, T. G. Walker, and M. Saffman,
Demonstration of a neutral atom controlled-NOT quantum gate,
Phys. Rev. Lett. \textbf{104}, 010503 (2010).


\bibitem{Levine2019}
H. Levine, A. Keesling, G. Semeghini, A. d Omran, T.-T. Wang, S. Ebadi, H. Bernien, M. Greiner, V. Vuleti\'{c}, H. Pichler, and M. D. Lukin,
Parallel implementation of high-fidelity multiqubit gates with neutral atoms,
Phys. Rev. Lett. \textbf{123}, 170503 (2019).

\bibitem{Zhaopeizi2017}
P.-Z. Zhao, X.-D. Cui, G.-F. Xu, E. Sj\"{o}qvist, and D.-M. Tong,
Rydberg-atom-based scheme of nonadiabatic geometric quantum computation,
Phys. Rev. A \textbf{95}, 052316 (2017).

\bibitem{Zhaopeizi2018}
P.-Z. Zhao, X. Wu, T.-H. Xing, G.-F. Xu, and D.-M. Tong,
Nonadiabatic holonomic quantum computation with Rydberg superatoms,
Phys. Rev. A \textbf{98}, 032313 (2018).

\bibitem{CPShen2019}
C.-P. Shen, J.-L. Wu, S.-L. Su, and E. Liang,
Construction of robust Rydberg controlled-phase gates,
Opt. Lett. \textbf{44}, 2036-2039 (2019).






\bibitem{xia2015}
T. Xia, M. Lichtman, K. Maller, A. W. Carr, M. J. Piotrowicz, L. Isenhower, and M. Saffman,
Randomized benchmarking of single-qubit gates in a 2D array of neutral-atom qubits,
Phys. Rev. Lett. \textbf{114}, 100503 (2015).

\bibitem{Saffman2016}
M. Saffman,
Quantum computing with atomic qubits and Rydberg interactions: Progress and Challenges,
J. Phys. B \textbf{49}, 202001 (2016).

\bibitem{Zhang2012}
X.-L. Zhang, A. T. Gill, L. Isenhower, T. G. Walker, and M. Saffman,
Fidelity of a Rydberg-blockade quantum gate from simulated quantum process tomography,
Phys. Rev. A \textbf{85}, 042310 (2012).

\bibitem{Petrosyan2017}
D. Petrosyan, F. Motzoi, M. Saffman, and K. M{\o}lmer,
High-fidelity Rydberg quantum gate via a two-atom dark state,
Phys. Rev. A \textbf{96}, 042306 (2017).

\bibitem{wujinlei2021}
J.-L. Wu, Y. Wang, J.-X. Han, Y.-Y Jiang, J. Song, Y. Xia, S.-L. Su, and W.-b. Li,
Systematic error tolerant multiqubit holonomic entangling gates, Phys. Rev. Appl. \textbf{16}, 064031 (2021).

\bibitem{MengLi2021}
M. Li, F.-Q. Guo, Z. Jin, L.-L. Yan, E.-J. Liang, and S.-L. Su,
Multiple-qubit controlled unitary quantum gate for Rydberg atoms using shortcut to adiabaticity
and optimized geometric quantum operations,
Phys. Rev. A \textbf{103}, 062607 (2021).






\end{thebibliography}
\end{document}